\documentclass[12pt]{article}

\usepackage{graphics,epsfig}
\usepackage{amsbsy}
\usepackage{amsfonts}
\usepackage{amsmath}
\usepackage{graphicx}

\def\id{\protect{{1 \kern-.28em {\rm l}}}}

\def \rI {{\rm I}}
\def \II {{\rm II}}
\def \III {{\rm III}}

\def\k{\kappa}

\def\p{{\partial}}
\def\nn{\nonumber}

\def\dalemb#1#2{{\vbox{\hrule height .#2pt
        \hbox{\vrule width.#2pt height#1pt \kern#1pt
                \vrule width.#2pt}
        \hrule height.#2pt}}}

\let\a=\alpha \let\b=\beta \let\g=\gamma \let\d=\delta \let\e=\epsilon
\let\z=\zeta  \let\th=\theta  \let\k=\kappa
\let\l=\lambda \let\m=\mu  \let\x=\xi \let\p=\pi 
\let\s=\sigma \let\t=\tau   \let\c=\chi 
\let\vp=\varphi \let\vep=\varepsilon
\let\w=\omega      \let\G=\Gamma \let\D=\Delta \let\Th=\Theta \let\L=\Lambda
 \let\P=\Pi \let\S=\Sigma  
\let\C=\Chi \let\W=\Omega
\let\la=\label \let\ci=\cite 
  
\def\nn{\nonumber} \def\bd{\begin{document}} \def\ed{\end{document}}
\def\ds{\documentstyle} \let\fr=\frac \let\bl=\bigl \let\br=\bigr
\let\Br=\Bigr \let\Bl=\Bigl
\let\bm=\bibitem
\let\na=\nabla
\def\tU{{\widetilde U}}
\let\pa=\partial \let\ov=\overline
\def\ie{{\it i.e.\ }}
\newcommand{\be}{\begin{equation}}
\newcommand{\ee}{\end{equation}}
\def\ba{\begin{array}}
\def\ea{\end{array}}
\def\ft#1#2{{\textstyle{{\scriptstyle #1}\over {\scriptstyle #2}}}}
\def\fft#1#2{{#1 \over #2}}
\def\F#1#2{{ F_{#1}^{(#2)} }}
\def\cF#1#2{{ {\cal F}_{#1}^{(#2)} }}

\def\={\, =\, }
\def\+{\, +\, }
\def\-{\, -\, }

\def\R{{\bf R}}
\def\sst#1{{\scriptscriptstyle #1}}
\def\oneone{\rlap 1\mkern4mu{\rm l}}
\def\e7{E_{7(+7)}}
\def\td{\tilde}
\def\wtd{\widetilde}
\def\im{{\rm i}}
\newcommand{\ho}[1]{$\, ^{#1}$}
\newcommand{\hoch}[1]{$\, ^{#1}$}
\newcommand{\bea}{\begin{eqnarray}}
\newcommand{\eea}{\end{eqnarray}}
\newcommand{\ra}{\rightarrow}
\newcommand{\lra}{\longrightarrow}
\newcommand{\Lra}{\Leftrightarrow}
\newcommand{\ap}{\alpha^\prime}
\newcommand{\bp}{\tilde \beta^\prime}
\newcommand{\cB}{{\cal B}}
\newcommand{\cO}{{\cal O}}
\newcommand{\vecx}{\vec{x}}
\newcommand{\vecy}{\vec{y}}
\newcommand{\vecp}{\vec{p}}
\newcommand{\vecq}{\vec{q}}
\newcommand{\tr}{{\rm tr} }
\newcommand{\Tr}{{\rm Tr} }

\newcommand{\cL}{{\cal L}}
\newcommand{\cA}{{\cal A}}
\newcommand{\cD}{{\cal D}}
\def\sst#1{{\scriptscriptstyle #1}}

\def\ve{\varepsilon}
\def\vf{\varphi}
\def\F{\Phi}
\def\wg{\wedge}

\def \foot {\footnote}
\def \bi{\bibitem}

\def \tr {{\rm tr}}
\def \ha {{1 \over 2}}
\def \td {\tilde}
\def \ci{\cite}
\def \N {{\mathcal N}}
\def \ww {\Omega}
\def \const {{\rm const}}
\def \ss {\sum_{i=1}^3 }
\def \t {\tau}
\def\S{{\mathcal S} }
\def \nn {\nu}
\def \XX {{\rm X}}

\def \lra {\leftrightarrow}
\def \vom {{\bar \omega}}
\def \E {{\mathcal  E}} \def \J {{\mathcal  J}}
\def \YY {{\rm Y}}

\def \d {\del}
\def \rJ {{J}}
\def \sms {sigma models\ }
\def \sm {sigma model\ }
\def \L {\Lambda}
\def \gl {\ell}
\def \tr {{\rm tr\ }}
\def\z{\zeta}
\def\zi{\zeta_1}
\def\zii{\zeta_2}
\def\K{\mbox{K}}
\def\eE{\mbox{E}}   \def \vt {\vartheta}
\def \vr {\varrho}
\def \wup {w}

\def\dg{\dagger}
\def\a{\alpha}
\def\b{\beta}
\def\e{\varepsilon}
\def\p{\phi}
\def\ap{\alpha^\prime}
\def\I{{\cal I}}

\def\R{{\bf R}}
\def\Z{{\bf Z}}
\def\C{{\bf C}}
\def\P{{\bf P}}
\def\xb{{\bar X}}
\def\Tr{{\rm  Tr}}
\def\tr{{\rm  tr}}

\def \del{\partial}
\def \a {\alpha}
\def \aa {{\a'}}
\def\g{\gamma}
\def\s{\sigma}
\def\z{\zeta}
\def\zi{\zeta_1}
\def\zii{\zeta_2}
\def\ov{\over}

\def\I{{\cal I}}
\def\J{{\mathcal J}}
\def \ok {{1\ov \k}}
\def\LL{{\mathcal L }}
\def \jL {{J}}
\def \om {\omega}
\def \cL {{\mathcal L}} \def \cH {{\mathcal H}}
\def\E{{\mathcal E}}
\def\w{\omega}
\def\b{\beta}
\def\l{\lambda}
\def\eps{\epsilon}
\def\vep{\varepsilon}
\def \De {{\mathcal D}}

 \def \cV {{\cal V}}
\def  \Jt {  {J}_{\rm tot}    }

\def \k {\kappa}
\def\foot{\footnote}
\def \four{{\textstyle {1\ov 4}}}
 \def \third { \textstyle {1\ov 3
}}
\def\det{\hbox{det}}
\def \ci {\cite}

\def \foot {\footnote}
\def \bi{\bibitem}

\def \tr {{\rm tr}}
\def \ha {{1 \over 2}}
\def \tid {\tilde}
\def \vv {{\rm v}}
\def \tl {{\tilde \l}}
\def \XX {{\rm X}}
\def \ta {{\tilde \a}}
\def \fo { {1\ov 4}}
\def \ep {\epsilon}
\def \inti {{\int^{2\pi}_0 {d \sigma \ov 2 \pi}}}

\def \d {\partial}
\def \K {{\rm S}}
\def \el {\ell}
\def \Tr {{\rm Tr}}
\def \P {\Phi}
\def \l  {\lambda}
\def \tl {{\tilde \l}}
\def \bl {{\tilde \l}}
\def \const {{\rm const}}
\def \V {v}

\def \bv {v^*}
\def \vv {{\rm v}}
\def \LL {{\mathcal L}}
\newcommand{\PV}[1]{P_{\!\!_{V_{#1}}}}

\def \bL {\ell}
\def \M {{\mathcal M}}
\def \N {{\mathcal N}}
\def \S {{\rm S}}
\def \vn {\vec n}
\def \tl {\td \l}
\def \td {\tilde}
\def \Prod {\Pi}
\def \O {{\mathcal O}}
\def \Q {{\rm  Q}}
\def \D {\Delta}
\def \N {{\mathcal N}}
\def\tN{{\tilde N}}

\def \m {\mu}
\def \vs {\vec \s}
\def \ie {i.e.}

\def \cD {{\cal D}}

\def  \le  {\l_{\rm eff}}

\def \rS {{\rm S}}
\def\as{{\a}}
\newcommand{\bra}[1]{\mbox{$\langle #1 |$}}
\newcommand{\ket}[1]{\mbox{$| #1 \rangle$}}

\newcommand{\auth}{AUTHORS}

\def\thb{\bar{\theta}}
\def\Thb{\bar{\Theta}}
\def\barp{\bar{p}}
\def\barq{\bar{q}}
\def\barc{\bar{c}}
\def\bard{\bar{d}}
\def\e{\epsilon}

\def \bi{\bibitem}
\def \la {\label}

\def \l {\lambda}
\def\foot{\footnote}
\def \tl  {{\tilde \l}}
\def \sql {{\sqrt \l}}
\def \adss {$AdS_5 \times S^5$\ }
\newcommand{\rf}[1]{(\ref{#1})}
\def \ov {\over}

\def\th{\theta}
\def\Th{\Theta}
\def\vth{\vartheta}

\def\vth{\vartheta}

\def\ra{\rightarrow}
\def\N{{\cal N}}
\def\F{{\cal F}}
\def\cc{\circ}
\def\eqv{\equiv}

\def\ni{\noindent}
\def \ha{{1\ov 2}}
\def \bw {{\rm w}}

\def\r{{\rm r}}

\def \cT {{\cal T}}
\def \no {\nonumber}

\def \J {\mathcal{J}}
\def \del {\partial}

\def \bps {{\bar \psi}}
\def \sqbl {\sqrt{\bar \lambda}}

\def\dF{\dot{F}}
\def\dG{\dot{G}}
\def\df{\dot{f}}
\def \E {{\cal E}}

\def \S {{\cal S}}
\def \J {{\cal J}}

\def\ms{\mathcal{S}}
\def\mj{\mathcal{J}}
\def\soj{\fr{\ms}{\mj}}
\def \R {{\bf R}}
\def \om {\omega}
\def \tH {\widetilde H}
\def \bE {\bar E}
\def \x {{\cal X}}

\def \hV {{\hat V}}

 \def \bb {\bar \beta}
\def \W {{\cal E}}

\def \bi{\bibitem}
\def \la {\label}

\def \l {\lambda}
\def\foot{\footnote}
\def \tl  {{\tilde \l}}
\def \sql {{\sqrt \l}}
\def \sqtl {{\sqrt {\tilde \l}}}

\def \HH {{\rm E}}
\def \cS {{\cal S}}
\def \cL {{\cal L}}

\def \adss {$AdS_5 \times S^5$\ }

\def \D {\Delta}
\def \thet {\theta}
 \def \t {\tau}
 \def \p {\phi}
 \def \r {\rho}
 \def \rN {{\rm N}}
 \def\tw{{\tilde w}}
 \def\hJ{{J}}
 \def\hw{{w}}
 \def\hl{{\lambda}}
 \def\hth{{\theta}}
 \def\NN{{\cal N}}
 \def \bv {{ \bar w}}
\def \vn {{\vec n}}
\newcommand{\sfrac}[2]{{\textstyle\frac{#1}{#2}}}
\def \bl {{ \bar \lambda}}
\def \bp {{\bar p}}
\def \bu {{\bar u}}
\def \sha {\sfrac{1}{2}}
\def \w {\omega}
\def \ov {\over}
\def \vl { \vec \ell}
\def \varpi {{\rm w}}
\def \OO {{\cal O}}
\def \bG {\bar \G}

\def \c {\gamma}

\def \ss {{\rm s}}

\def \ve {\varepsilon}
\def \pa{\partial}
\def \I {{\cal I}}
\def \LL {{\cal L}}
\def \ep {\epsilon}
\def \R {{\rm R}}
\def \tilt {{\tilde t}}

\def\pic #1#2{\hbox{\lower#1pt\hbox{~\mbox{\epsfxsize=20truemm \epsffile{#2}}}}}

\def\pic #1#2#3{\hbox{\lower#1pt\hbox{~\mbox{\includegraphics[scale=#3]{#2}}}}}

\def \bt {\bar\theta}
\def \te {\theta}

\def \cc {{\rm f}}
\def \d {\delta}

\def \cL {{\cal L}}
\def \S  {{\cal S}}
\def \pp {{q}}
\def \vt {\vartheta}
\def \mm {{\cal  \ell}}
\def \Z {{\cal Z}}
\def \pa {\partial}
\def \C {{\cal C}}
\def \be {\bea}
\def \ee {\eea}
\def \c {\gamma}  \def \d {\delta}
\def \eps {\epsilon}

\def \bp {\begin{pmatrix}}  \def \ep {\end{pmatrix}}

 \def \T {{\cal T}}

\def \vr {\varrho}\def
\vt {\vartheta}
\def \F {{_F}}
\def \DD {{\rm D}}

\def \bp {\begin{pmatrix}}  \def \epm {\end{pmatrix}}
\def \ha {{\textstyle{1 \ov 2}}}

\begin{document}
\overfullrule=0pt
\parskip=2pt
\parindent=12pt
\headheight=0in \headsep=0in \topmargin=0in \oddsidemargin=0in

\vspace{ -3cm} \thispagestyle{empty} \vspace{-1cm}
\begin{flushright} Imperial-TP-AT-2008-3
\end{flushright}
\begin{center}
 \vspace{2cm}
{\Large\bf
 Quantum  corrections to energy        \\
\vspace{0.3cm}
of short spinning string in $AdS_5$
 }

 \vspace{.5cm} {
  A. Tirziu$^{a,}$\footnote{atirziu@purdue.edu}
 and A.A.
 Tseytlin$^{b,}$\footnote{Also at
 Lebedev  Institute, Moscow.\ \
  tseytlin@imperial.ac.uk
 }}\\
 \vskip 0.3cm

{\em
$^{a}$Department of Physics, Purdue  University,\\
W. Lafayette, IN 47907-2036, USA.\\
\vskip 0.08cm
$^{b}$  The Blackett Laboratory, Imperial College,
London SW7 2AZ, U.K. }

\end{center}

 \begin{abstract}
Motivated by a desire to shed light on the
strong coupling  behaviour of dimensions of ``short''
gauge theory operators
we consider the famous example of
 folded spinning string  in $AdS_5$
in the limit of small semiclassical spin parameter
$\S = { S \ov \sql}$. In this limit  the string  becomes short
and is  moving in a near-flat central region of $AdS_5$.
Its energy scales with spin as $E = \lambda^{1/4} \sqrt{2 S}\ [ a_0 +
a_1 S + a_2 S^2 + ...]$. We explicitly
compute the leading 1-loop quantum
\adss superstring corrections  to the short-string limit coefficients $a_0$ and $a_1$
and show, in particular,  that
$a_1$ receives a  contribution containing $\zeta(3)$.

\end{abstract}
\newpage

\renewcommand{\theequation}{1.\arabic{equation}}
 \setcounter{equation}{0}

\setcounter{equation}{0} \setcounter{footnote}{0}
\setcounter{section}{0}

\section{Introduction}

The remarkable  progress achieved recently in uncovering the integrable structure
underlying the
spectrum of planar $\N=4$  SYM theory or  the free \adss superstring theory
was largely limited to a sector of gauge theory operators with large number of
fields/derivatives
or strings with large values of  quantum numbers like spins.
It is  important  to try to learn more about dimensions/energies of ``short''
operators/strings  and a step in that direction is to  study
 quantum corrections to energies
of  strings carrying parametrically small values of spins.

With this   motivation in mind here we revisit the computation of the 1-loop
quantum correction to the energy of the prototypical example
of rotating string -- folded rotating string  located at the center of
$AdS_5$ \ci{dev,gkp}.

The classical  energy of this string is proportional to string tension, i.e.
$E_0 = \sql\ \E( \S), \ \  \S= { S \ov \sql} $  and in the limit of large
 $\S$ one finds \ci{gkp}:
$E_0 = S + { \sql \ov \pi} \ln S + ... $.
In general, the  radial coordinate $\r$ of the global $AdS_5$ space
($ds^2 = - \cosh^2 \r\ dt^2 + d\r^2 + \sinh^2 \r \ d\Omega_3^2$)  is
expressed in terms of  an elliptic function of the spatial string coordinate
$\s$  and thus finding the explicit   form of  the 1-loop
correction \ci{ft1} to the energy $E_1$
of this soliton solution of 2d  string sigma model
appears to be  technically challenging.
The analytic form of the quantum correction can be found
 in the limit of large $\S$  when
the ends of the string
reach the boundary of the $AdS_5$. Then the  solution drastically
simplifies ($\rho$ becomes
linear in $\s$)  \ci{ft1,ftt}  and   one finds   that
 $E_1= c_1 \ln S + ..., \ \  c_1= - {3 \ln 2\ov \pi}$.

Since rotation of the string balances the contracting  effect of its tension,
 smaller values of the spin correspond to smaller values of the
 length of the string
whose  center of mass is at $\r=0$:\  $\S$ essentially measures the
length of the string.
 Since the $AdS_5$  space is nearly flat at  the vicinity
of $\r=0$, the slowly rotating   (i.e. small) string
with $\S \ll 1$    should  have essentially the same classical
 energy as in flat space \ci{gkp}, i.e.
$E_0 = \sqrt{ 2 \sql S} +...$.

Below we shall  expand   the general expression  for the 1-loop correction to
the energy of the spinning
string in \ci{ft1} (given  by a sum of logarithms of determinants
of the 2d second order differential operators depending on the string background)
in the ``short string'' limit  $\S \ll 1$  and   find explicitly
 the coefficients
of the first two leading terms in the  small spin  expansion
of the 1-loop  energy.

Our results can be summarized as follows. Given  the energy  $E(S, \l) $ of the corresponding
state in the AdS/CFT spectrum we may  expand it at large $\l$  with $\S = { S \ov \sql}$
fixed, i.e. in the semiclassical string limit. Expanding {\it then} in  the limit $\S \ll 1$, i.e.
$S \ll \sql $,  and re-expressing $E$ as a function of $S$ and $\l$  one is to find
\bea
&& E(S,\lambda)=
 \lambda^{{1}/{4}}\sqrt{2 S}\ \Big[h_0(\lambda) + h_1(\lambda)S + h_2(\lambda) S^2 + ...\Big]\ ,
\    \label{lmt} \\
&&  h_n =  \frac{1}{(\sqrt{\lambda})^n}( a_{n0} + \frac{a_{n1} }
{\sqrt{\lambda}}+\frac{a_{n2}}{(\sql)^2}+...) \ ,
\  \ \ \ \  \   \l \gg 1, \ \   {S \ov \sql}={\rm fixed} \ll 1 \ . \la{ha}
\eea
In the classical   string theory limit
\be \la{cl}
a_{00}=1\  , \ \ \ \ \  \ \ a_{10}= { 3 \ov 8}  \ , \ \ \ \ \ \ \ a_{20}=-{21\ov 128}  \ , ... \ee
while our 1-loop string computation gives
\be \la{on}
a_{01}=3-4 \ln 2 \approx 0.227\  , \ \ \ \ \  \ \
 a_{11}=-\frac{1219}{576}  + { 3 \ov 2}  \ln 2 +  { 3 \ov 4} \zeta (3)   \approx -0.175
 \ .   \ee
 The leading $\sqrt{2 S}$   term has  the same form as in
the flat-space string  theory, but its  coefficient  gets renormalized
from its classical value 1.
Classically, a short string in the middle of $AdS_5$   does not feel the curvature
so its energy is   the same as in flat space.
In flat space the string fluctuations
are essentially quadratic  and massless  (as happens in  the in light-cone gauge). They
thus decouple  from the rotating string
background (as we shall  discuss explicitly  for the Green-Schwarz  string  in covariant gauge
in  Appendix A below)  and  do not  change the  classical
$E= \sqrt{ 4 \pi T S}$   expression  (here $T= {1 \ov 2 \pi \a'}$).
In curved space  the bosonic fluctuations feel  the curvature and as a result get mass
depending on the string background; the fermionic fluctuations  get similar mass due to
their coupling to the RR  5-form background. While most of the resulting  contributions  to the leading
$\sqrt{2S}$ term in the energy cancel between the  bosonic  and fermionic terms,
there is a nontrivial residue (proportional to the $\sigma$-derivative
of the fermionic  mass term)\foot{This contribution
   was missed in the original version of this paper.}
  leading to the non-zero value of the 1-loop  $a_{01}$  coefficient.

 Explicitly,
 \rf{lmt} can be written  as
\bea
 E(S,\lambda)=
 \lambda^{{1}/{4}}\sqrt{2 S}\ \Big[(1  +
  \frac{a_{01}}{\sqrt{\lambda}} + ...)     +  (a_{10}   + \frac{a_{11}}{\sqrt{\lambda}} + ...) { S \ov \sql }     +
  (a_{20}   + \frac{a_{21}}{\sqrt{\lambda}} + ...) { S^2 \ov (\sql)^2 }   +... \Big]
  \label{mt}  \ee
In contrast to the large spin (or ``long string'')  limit   where
the limits of large $\l$ and large $S$ appear to commute\foot{The
perturbative string theory  and perturbative gauge theory limits are
actually different as limits of functions on
 the two-parameter space $(\l, S)$:
 in string theory  one assumes  $\l \gg 1$
with $\S= {S \ov \sql}$ fixed and then takes $\S$ large;
in gauge theory one assumes  $\l \ll 1$
with $\S$ fixed and then takes $S$ large.  However, this
appears not to matter for the leading $\ln S$ term
which can be described by a single universal interpolating function of $\l$
(cusp anomaly).}
(and thus one finds the same
$S$ dependence  of the gauge theory anomalous
dimension and string theory energy
at both  weak  and strong coupling,
 $E=S + f(\l) \ln S + ...$, with  $f(\l \ll 1 )= c_1 \l + c_2 \l^2 + ..., \
\ f(\l \gg1 ) = \sql ( b_0 + {b_1 \ov \sql} + ...)$)
here one cannot  directly continue  \rf{lmt} to small $\l$ and small $S$.

Indeed, the  anomalous   dimensions  of low-twist  gauge-theory
operators like $\tr ( \Phi   D_+^S \Phi)$  computed
for small $\l$ and fixed $S$  (see, e.g.,
\ci{kot})  and then formally expanded in small $S$ limit  scale  as\footnote{If the twist two operator in question is assumed to be
from the sl(2)  sector then
it is   BPS for $S=0$ so  that   $q_0 =2$.
}
\begin{equation}
E( \lambda, S) = q_0(\lambda)+ q_1(\lambda)  S + q_2(\lambda) S^2 +  O(S^3) \ , \ \ \ \ \ \ \ \        \l \ll 1, \ \
  S={\rm fixed}    \ ,   \label{ak}
\end{equation}
where
\begin{equation} \la{uy} q_n(\lambda)= d_{n0} +  d_{n1}  \lambda + d_{n2}  \lambda^ 2
+ ... \ , \ \ \ \ \ \ \ \  \ \ \ \  n=0,1,2,...\    \  .
\end{equation}
To relate the  ``small-spin''  string theory  \rf{lmt}  and the gauge theory
\rf{ak} expansions
one would need to resum the series in both arguments $(\l,S)$,
e.g.,
  first sum up
the weak-coupling expansion in \rf{ak} and then re-expand the
 result first in large  $\l$
for fixed $\S={S\ov \sql} $ and then in small $\S$.

In view of the  need for  this resummation which is, in fact,  a
generic situation in comparing the semiclassical string theory and the
perturbative gauge theory expansions\foot{Analogous resummation
is needed to compare the weak coupling gauge theory expansion
for anomalous dimensions of sl(2) sector operators
in the limit $\l \ll 1$ with $  \ J \gg 1, \ S \gg 1$, \
$j= { J \ov \ln S}$=fixed and $ j < 1$   with
the strong-coupling string theory expansion in the limit
$\l \gg 1$  with $   \J = { J \ov \sql}, \ \S= {S \ov \sql }$, \
$ \ell \equiv { \J\ov  \ln \S}$=fixed and $\ell   < 1$
(see \ci{ft1,bfst,bgk,ftt,am,rt,frs}).}
it is not   clear at the moment  how to directly interpret
  our result \rf{mt}
as a strong coupling limit  of a gauge-theory anomalous dimension.

\

We shall start  in section 2 with a review of the folded spinning
string solution and its small spin expansion \ci{gkp}.

In section 3 we shall first  recall the general expression
for the quadratic fluctuation Lagrangian $\td L$ \ci{ft1}
of the \adss superstring \ci{mt}  near the folded spinning
string solution.  We will then expand
the coefficients in $\td L$   in the small spin or short string
parameter $\eps= \sqrt { 2 \S} + ...$. This expansion  may be viewed
as a  particular case of a near flat space expansion  of the quantum \adss
superstring. We will then compute
the
leading $O(\eps)$  term in the 1-loop  string energy
determining  the coefficient $a_{01}$ in \rf{on},\rf{mt}.

In section 4 we shall expand the  2d determinants  that enter the
expression  for the 1-loop partition function to first two leading
orders in $\eps$ and compute the
value of the coefficient $a_{11}$  in \rf{on},\rf{mt}.

In Appendix A   we shall present the flat-space  Green-Schwarz string
 analog of this computation  showing explicitly (in a covariant $\kappa$-symmetry gauge)
 why the classical  $E= \sqrt{ 4 \pi T S}$
 expression is not renormalized by quantum fluctuations.

In Appendix B we shall briefly discuss  how to
generalise  our computation to the case of the short string expansion of the
folded spinning string solution which also carries a momentum $J$
 in $S^5$ \ci{ft1} (details of this  case  are worked in the follow-up paper \ci{BT}).

In Appendix C we shall mention
 a curious regularization scheme
  ambiguity  which appears, in particular,
 when interchanging  a sum with an integral in certain 1-loop  terms.

\renewcommand{\theequation}{2.\arabic{equation}}
 \setcounter{equation}{0}

\setcounter{equation}{0} \setcounter{footnote}{0}

\section{Short string limit  of   folded spinning string solution}

Let us start with a review of the classical solution for the
folded  string spinning in the $AdS_3$ part of $AdS_5$,
\begin{equation}\la{so}
 t= \kappa \tau, \quad \phi= w \tau, \quad \rho=\rho(\sigma) \ , \ \ \ \ \ \ \
ds^2= -\cosh^2 \rho\ dt^2 + d \rho^2 + \sinh^2 \rho\ d \phi^2
\ , \end{equation}
where
\begin{equation}
\rho'^2 = \kappa^2 \cosh^2 \rho - w^2 \sinh^2 \rho \ .   \label{snh}
\end{equation}
$\rho$ varies from $0$ to its maximal  value $\rho_*$
\begin{equation}
\coth^2 \rho_* = \frac{w^2}{\kappa^2}\equiv 1+ \frac{1}{\epsilon^2} \ .
\end{equation}
Thus $\eps$ measures the length of the string.
The solution of the  differential equation (\ref{snh}), i.e.
\begin{equation}
\rho' = \pm \kappa \sqrt{1-\eps^{-2} {\sinh^2 \rho}}\ , \quad \quad \rho(0)=0
\end{equation}
 can be written in terms of the Jacobi function ${\rm sn}$
\begin{equation}
\sinh \rho=\epsilon \ {\rm sn}({\kappa\eps^{-1}  \sigma},\ -\epsilon^2) \ . \label{mlq}
\end{equation}
The periodicity in $\sigma$ implies the following condition on the parameters \ci{gkp}
\begin{equation}
\kappa=\epsilon \  _2 F_1(\frac{1}{2},\frac{1}{2};1;-\epsilon^2) \la{uu}  \ .
\end{equation}
The classical energy $E_0=\sqrt{\lambda} \mathcal{E}_0$ and the spin $S=\sqrt{\lambda}\mathcal{S}$ are found to be
\begin{equation}
\mathcal{E}_0=\epsilon  \ _2 F_1(-\frac{1}{2},\frac{1}{2};1;-\epsilon^2),
\ \ \  \qquad \mathcal{S}=
\frac{\epsilon^2}{2}\sqrt{1+{\epsilon^2}} \ _2F_1(\frac{1}{2},\frac{3}{2};2;-\epsilon^2)  \label{qdr}
\end{equation}
Here we will
be   interested
 in the short string limit $0 < \epsilon \ll 1$   in which
\begin{equation}
\rho_*=\epsilon   - {1 \ov 6}\epsilon^3 +  O(\epsilon^{5})  \ .
\end{equation}
In the strict limit
$\epsilon=0$  or $\k=0$ we get   $\rho=\rho_*=0$,  so that
the  string  shrinks to a point   with $E=0$.\foot{Note that in this
 limit the string disappears
instead of reducing to a  massless point particle  with  non-zero momentum
 moving along null geodesic. This corresponds in flat space to considering
a massive string state in the rest  frame  (which is possible in
covariant quantization).
In contrast to the flat space case where adding a non-zero center of
 mass momentum  can be achieved by a Lorentz boost,
adding  a  motion of the spinning string
 center of mass in curved  $AdS_5\times S^5$ space is a nontrivial
operation (different parts of the string move along different geodesics)
which leads in general to a new nontrivial configuration.}

From  (\ref{qdr}) we obtain  in the small $\epsilon$ or
the small  $\mathcal{S}$ limit
\bea
\epsilon=\sqrt{2 \mathcal{S}}-\frac{1}{4 \sqrt{2}}\mathcal{S}^{{3}/{2}}+... \ ,
\ \ \ \ \ \ \ \
\ \ \ \mathcal{E}_0=\sqrt{2 \mathcal{S}}+\frac{3}{4 \sqrt{2}}\mathcal{S}^{{3}/{2}}+...
\label{mfj} \ ,
\eea
so the short string limit corresponds to $\S  \ll 1$ and
 the expansion of the  energy looks like
\begin{equation}\la{fj}
E_0(S,\lambda)= \lambda^{{{1}/{4}}}\sqrt{2 S}+\frac{3}{4 \sqrt{2}}\lambda^{{-{1}/{4}}} S^{{3}/{2}}+O(S^{5/2}) \ .
\end{equation}
For the purpose of computing the $1$-loop correction to the energy
 to order $O(S^{{3}/{2}})$ we will  need the expression  for $\r(\s)$
 to order
  $\epsilon^4$. Expanding the exact solution  (\ref{mlq}) in powers of $\epsilon$ we obtain
\begin{equation}
\sinh \rho = \epsilon \sin \sigma - \frac{\epsilon^3}{4}\sin \sigma\ \cos^2 \sigma + O(\epsilon^5)  \label{abe}
\end{equation}
Other useful expansions  are
\bea
&&\kappa=\epsilon(1 -\frac{\epsilon^2}{4}+...) \ ,\ \ \  w=1 + \frac{\epsilon^2}{4} + ...\ ,
\ \   \quad \rho'=\epsilon \cos \sigma - \frac{\epsilon^3}{4}\cos^3 \sigma +...
 \ , \\
&&\kappa \sinh \rho=\epsilon^2 \sin \sigma -\frac{\epsilon^4}{8}(3+ \cos 2 \sigma)\sin \sigma+...\ ,\\
&& w \cosh \rho=  1+ \frac{\epsilon^2}{4}(1+ 2 \sin^2 \sigma)-\frac{\epsilon^4}{64}(8-\cos 4 \sigma)+... \ .
\eea
The above small spin expansion is an example of
 a near flat space expansion: the
 leading-order in $\eps$  solution
can  be identified with the  folded  spinning  string solution in the flat space
\begin{equation}
t=\eps  \tau\ , \quad \rho=\eps  \sin \sigma\ , \quad \phi=\tau \ , \ \ \ \ \ \
ds^2=-dt^2 + d \rho^2 +\rho^2 d \phi^2 \ ,
\end{equation}
where $\eps$ is an arbitrary  constant amplitude.
The energy and the spin  then satisfy the usual flat-space  Regge relation
(we use  string  tension $T= {\sql \ov 2 \pi}$)
\begin{equation}
E_0 = \eps \sqrt{\lambda}\ , \qquad S=   \frac{\eps^2}{2}\sqrt{\lambda} \ ,
\ \ \ \ \ {\rm i.e.} \ \ \ \ \ \
\E_0 = \lambda^{{{1}/{4}}} \sqrt{2 S}  \ . \label{mbj}
\end{equation}
In the flat space case  this is the  exact expression for any value of $S$ (cf. \rf{fj})
which also does not receive quantum corrections (see Appendix A).

\renewcommand{\theequation}{3.\arabic{equation}}
 \setcounter{equation}{0}

\setcounter{equation}{0} \setcounter{footnote}{0}

\section{$1$- loop correction to $\sqrt{S}$ term in short string  energy }

Following \ci{ft1} and expanding the \adss  string action \ci{mt}
in conformal gauge  to
quadratic order in fluctuations near the folded spinning string
  one finds
$
\td S=-\frac{\sqrt{\lambda}}{4 \pi}\int d \tau \int_0^{2 \pi} d \sigma\ \td {L}
$  with the bosonic part $(a=0,1)$
\begin{eqnarray}
\td {L}_B&=& - \partial_a \td {t} \partial^a \td {t}- \mu_t^2 \td {t}^2 +   \partial_a \td {\phi} \partial^a \td {\phi}+ \mu_{\phi}^2 \td {\phi}^2\nonumber\\
&+& 4 \td {\rho} (\kappa \sinh \rho\ \partial_0 \td {t} - w \cosh \rho\ \partial_0 \td {\phi})+  \partial_a \td {\rho} \partial^a \td {\rho}+\mu_{\rho}^2 \td {\rho}^2\nonumber\\
&+& \partial_a {\beta}_u \partial^a {\beta}_u +\mu_{\beta}^2 {\beta}_u^2 +
 \partial_a {\varphi} \partial^a {\varphi}+\partial_a {\chi}_s \partial^a {\chi}_s \ ,   \label{lag}
\end{eqnarray}
where
\begin{equation}
\mu_t^2= 2 \rho'^2 -\kappa^2, \ \ \quad \mu^2_{\phi}=2 \rho'^2 -w^2, \ \
\quad \mu^2_{\rho}=2 \rho'^2 -w^2-\kappa^2,\ \
\quad \mu_{\beta}^2=2 \rho'^2  .    \label{as}
\end{equation}
Here $\beta_u$ ($u=1,2$) are two $AdS_5$ fluctuations
transverse to the $AdS_3$ subspace in which the string is moving, while
 $\varphi,\chi_s$ ($s=1,2,3,4$) are fluctuations in $S^5$.

 The fermionic part of the quadratic fluctuation Lagrangian   can be put into the form \ci{ft1}
 \be \la{feq}
 \td {L}_F=   2   \bar  \vt  \DD_\F   \vt \ , \ \ \ \ \ \ \ \
 \DD_\F=  i( \G^a \del_a   - \mu_\F \G_{234}) \  , \ \ \ \  \ \ \ \
 \mu_\F= \rho'  \   , \ee
 and  can be interpreted as describing a system of 4+4   2d Majorana  fermions with $\s$-dependent mass
 $\m_F$.
 Let us briefly recall the derivation \ci{ft1} of this expression.
 One starts with the quadratic fermionic  term  in the \adss  action
 and fixing the  conformal gauge $\sqrt{-g} g^{ab} = \eta^{ab}$
 and the $\kappa$-symmetry    gauge $\theta^1= \theta^2=\theta$
 (where  $\theta$ is a MW  10-d spinor) one    gets
 \be
 \td {L}_F=2 i \bar \theta \vr^a (D_a  - {i \ov 2}  \G_* \vr_a)  \theta \ , \ \ \ \ \ \ \ \
 \G_* = i \G_{01234}  \ ,     \ee
where $D_a = \del_a + { 1 \ov 4} \del_a X^M \omega^{AB}_M  \G_{AB} $,
\  $ \vr_a= \del_a X^M  E^A_M \G_A$.
Identifying $(t, \r, \phi)$ with the  directions $M=0,1,2$  and  introducing $\vt$ as
\be
\vt= \sqrt{ |\rho'| } \ e^{- {1\ov 2} \alpha  \G_{02} }\ \theta \ , \ \ \ \ \ \ \ \ \ \ \
\cosh \alpha = {\k \cosh \r \ov  |\r'|} \ , \ee
one ends up with \rf{feq}  where $\G^a= \eta^{ab} \G_b = ( - \G_0, \G_1)$
and $\bar \vt = \vt^T \G^0$  with $\Psi$  being real 16-component  (Weyl) spinor.\foot{For a discussion 
of spinor
notation see, e.g.,
Appendix A in \ci{rtt}.} Note that  since $\rho(\sigma)$ is periodic function, the same applies 
to $\alpha$, i.e.  the rotated fermions are periodic  in $\sigma$ just like the original ones. 
Note also  that  the fermionic mass term has its origin
in the RR 5-form coupling term in the  quadratic fermionic Lagrangian \ci{mt}.

Since  $\ln \det ( \G^0  \DD_\F) = \ln \det ( \DD_\F) = \ha \ln  \det (\DD_\F)^2 $
we conclude that  the fermionic  contribution to the 1-loop string partition function is determined
by the following second-order differential operator
\be \la{seco}
\Delta_\F \equiv (\DD_\F)^2 = - \del^a \del_a   + \mu'_\F \G_{1234}   + \mu^2 _\F \ ,  \ee
where  we used that $\G_{(a}\G_{b )} =\eta_{ab}, \  \{\G_a, \G_{234} \}=0 $, \ $ (\G_{234})^2 =-1$
and  that  $\mu_\F$   depends only on $\s$\  ($\mu'_\F\equiv  \del_1 \mu_\F$).
Furthermore, since  $(\G_{1234})^2 =1$ we can diagonalize this operator  so we will end up with
the following  contribution to the 1-loop 
2d effective action coming from 
4+4=8 effective fermionic degrees of freedom:\foot{This
 structure of
the fermionic contribution was  understood  in collaboration with M. Beccaria.}
\be   - \ha \Big(  4 \ln \det \Delta_{\F+} +  4 \ln \det \Delta_{\F-} \Big) \ , \la{fero}
\ee
where
\be \la{sep}
\Delta_{\F\pm}  = - \del^a \del_a   + \hat \m^2_{\F \pm} \ , \ \ \ \ \ \ \ \
 \hat \m^2_{\F \pm}\equiv \pm  \mu'_\F    + \mu^2 _\F=  \pm \r'' + \r'^2  \ .  \ee
Next, we
 expand the coefficients in the  fluctuation Lagrangian
in $\epsilon$ as discussed
 in the previous section. To leading order in $\eps$ we get
\bea
&& \mu_t^2= \epsilon^2 \cos 2 \sigma+ ..., \qquad
\mu^2_{\phi}=-1+( \cos 2 \sigma+\frac{1}{2})\epsilon^2+ ..., \\
&&  \mu^2_{\rho}=-1+( \cos 2 \sigma-\frac{1}{2})\epsilon^2+ ...,
\qquad \mu_{\beta}^2= 2 \epsilon^2 \cos^2 \sigma + ... \ ,  \label{las}
\\
&&  \hat \mu^2_{\F\pm }=\mp  \epsilon \sin \s + \epsilon^2 \cos 2 \sigma+ ... \ , \la{fem} \\
&& 4 \td {\rho} (\kappa \sinh \rho\ \partial_0 \td {t} - w \cosh \rho\ \partial_0 \td {\phi})=
  \td {\rho}\big\{4\epsilon^2 \sin \sigma\ \partial_0 \td {t}-[4+
 \epsilon^2 (1+2 \sin^2 \sigma) ]\partial_0 \td {\phi}\big\} \ .
\eea
If we  set $\epsilon$ to zero we are back to the  flat space case (see Appendix A):  indeed, the
only two coupled modes that are not massless are then described by
\begin{eqnarray}
\td {L}_0=    \partial_a \td {\phi} \partial^a \td {\phi}- \td {\phi}^2
- 4 \td {\rho}  \partial_0 \td {\phi}+  \partial_a \td {\rho} \partial^a \td {\rho}- \td {\rho}^2 \ ,
\end{eqnarray}
which becomes the Lagrangian for
 two massless modes after  a  $\tau$-dependent rotation
\begin{equation}
\td {\rho}=\eta_1 \cos \tau + \eta_2 \sin \tau, \qquad \td {\phi}= -\eta_1 \sin \tau + \eta_2 \cos \tau \ .  \label{rot}
\end{equation}
If we perform this rotation also at order $\eps^2$   we get
 $\tilde{L}_B=\td {L}_0+ \epsilon^2 \td {L}_1 + O(\eps^4)$
where $\td {L}_0$ is the same as in flat space and
a nontrivial part of the subleading term is\foot{We shall not
use this $\tau$-dependent form of the fluctuation Lagrangian for explicit
computations below.}
\begin{eqnarray}
\tilde{L}_1&=&-\cos 2 \sigma\ \td {t}^2 + (\sin^2 \tau+ \cos^2 \sigma) \eta_1^2 + (\cos^2 \tau+ \cos^2 \sigma) \eta_2^2+ 2(\eta_1 \cos \tau+ \eta_2 \sin \tau)\dot{\td {t}}\sin \sigma \nonumber\\
&-&2 (\dot{\eta}_1 \cos \tau+\dot{\eta}_2 \sin \tau)\td {t} \sin \sigma + 2 (\eta_1 \sin \tau-\eta_2 \cos \tau) \td {t}\sin \sigma\nonumber\\
&-&\eta_1 \eta_2 \sin 2 \tau -  \eta_1 \dot{\eta}_2(1+ 2 \sin^2 \sigma)   \ . \label{jhm}
\end{eqnarray}
The  order $\epsilon$ contribution  coming from $\r''$ term in the
effective  fermionic mass in \rf{sep}   will cancel out  in the sum of the 4+4 fermionic contributions
but  there will be an additional $\epsilon^2$  term coming from  the  double insertion of this term.
Let us first ignore this extra $\epsilon^2$
contribution   coming from the presence  of the $\mu'_\F =\r''$ term   in the fermionic masses.
Then  one may   argue  on general grounds that the leading
$\eps^2$  part of 1-loop correction to string energy should vanish.
Indeed, then the  1-loop correction to string energy will look like
(assuming all propagators were diagonalized)\foot{To cancel the leading
flat space term, i.e. to ensure that the total number of effective degrees of freedom is zero,
one  of course is to include also the conformal gauge ghost  contribution.}
\begin{equation}
\G_1  = \ha   \sum_i (-1)^{n_i} \ln \frac{\det[\partial_0^2-\partial_1^2+ \epsilon^2 M^2_i]}{\det[\partial_0^2-\partial_1^2]}
\sim \epsilon^2 \int d \tau d \sigma\ {\rm  Tr }\sum_i (-1)^{n_i} M^2_i +O(\epsilon^4)\label{dtw} \ .
\end{equation}
Since $t= \kappa \tau, \ \kappa = \eps + ...$
the  1-loop correction to string energy  is given by
\be \la{loi}
E_1 = {\G_1 \ov \k \T}  \ , \ \ \ \ \ \   \qquad \T \equiv \int d \tau \to \infty   \ . \ee
In general,  $M^2_i$ may be  non-trivial matrices  which depend on $\tau,\sigma$.
 Let us  now  recall that  the 1-loop
 logarithmic UV divergencies in the \adss   superstring action expanded near
 an
 arbitrary string solution  manifestly  cancel in the conformal gauge \ci{dgt,ft1}.
 The nontrivial UV logarithmic divergencies  have as their coefficient
  precisely  the  sum of the mass squared terms in the r.h.s. of  \rf{dtw};\foot{In general, the
  additional $\pm \r''$ terms in the fermionic mass  contributions  of course cancel against each other.}
   it vanishes for a generic on-shell string background, thus implying
the absence of the  $\eps^2$ term
  in the 1-loop string partition function  (again, modulo the addional
  $\eps^2$ contribution   coming from $\r''$ part  that we are
   temporarily ignoring).

Let us now verify  this cancellation  by  direct computation.
For the  contribution of the $\beta_u$ fields we get
(rotating to euclidean time, $\tau \to i \tau$, and factorizing the infinite time interval $\T$)
\begin{equation}
\det[-\partial_1^2 - \partial_0^2 + 2 \epsilon^2 \cos^2 \sigma]=
\T
\int {d{\omega}\ov 2\pi}  \det [-\partial_1^2+\omega^2+2 \epsilon^2 \cos^2 \sigma] \ .
\end{equation}
We can now use perturbation theory in $\epsilon^2$, i.e.
\begin{equation} \la{ll}
\ln \frac{\det [A+ \epsilon^2 B]}{\det A}=\epsilon^2 Tr[ A^{-1} B]+O(\epsilon^4) \ .
\end{equation}
Then to order $\eps^2$  (here $\s \in (0,2\pi)$)
\begin{equation}
\ln \frac{\det[-\partial_1^2+ \omega^2 +
2 \epsilon^2 \cos^2 \sigma]}{\det[-\partial_1^2+ \omega^2 ]} \approx
\epsilon^2 \sum_{n} \frac{2}{n^2 +\omega^2}\int_0^{{2 \pi}}
\frac{d \sigma}{2 \pi}  \cos^2 \sigma =\epsilon^2 \sum_{n} \frac{1}{n^2 + \omega^2}\ .  \label{mfg}
\end{equation}
Similarly, the $\eps^2$  contribution of the fermionic modes coming from $\r'^2$ term in \rf{sep}
 is proportional to
\begin{equation}
\ln \frac{\det[-\partial_1^2+ \omega^2 +  \epsilon^2 \cos^2 \sigma]}{\det[-\partial_1^2+ \omega^2 ]}
 \approx \epsilon^2 \sum_n \frac{1}{n^2 +\omega^2}\int_0^{{2 \pi}} \frac{d \sigma}{2 \pi}
  \cos^2 \sigma =\frac{\epsilon^2}{2} \sum_n \frac{1}{n^2 + \omega^2}\ .  \label{mfg1}
\end{equation}
The nontrivial part of the  euclidean partition function contributing to the $\eps^2$ term
under consideration is\foot{Here we choose not to rotate $\td  t \to i \td  t$ to
 make  all fluctuations having physical norm
but this can be easily done at any stage
 of what follows; we shall assume this rotation in the
free (flat)  part of the partition function.}
\begin{equation}
\tilde Z=\frac{\det^{\frac{8}{2}}[-\partial_0^2-\partial_1^2+\epsilon^2 \cos^2 \sigma]\
 \det^{\frac{2}{2}}[-\partial_0^2-\partial_1^2]
}{\ \det^{\frac{2}{2}}[-\partial_0^2-\partial_1^2+2
 \epsilon^2 \cos^2 \sigma]\ \det^{\frac{5}{2}}[-\partial_0^2-\partial_1^2]\ \det^{\frac{1}{2}}Q}
\end{equation}
involves the operator $Q$  on the space of the three  mixed fluctuations
 $\td  \r, \td  \phi, \td  t$
  in \rf{lag}
\begin{eqnarray}
Q=
 \bp     \partial_0^2+\partial_1^2-\epsilon^2 \cos 2 \sigma & 0 & -2 i \epsilon^2 \sin \sigma \partial_0 \\
      0 & -\partial_0^2-\partial_1^2-1+\epsilon^2 (\frac{1}{2} + \cos 2 \sigma) & 2 i \partial_0+
       i \epsilon^2 (\ha + \sin^2 \sigma) \partial_0 \\
      2 i \epsilon^2 \sin \sigma \partial_0  & -2 i \partial_0 - i \epsilon^2 (\ha + \sin^2 \sigma)
      \partial_0 & -\partial_0^2-\partial_1^2-1- \epsilon^2 (\frac{1}{2}-\cos 2 \sigma) \ep
\no
\end{eqnarray}
Since there is no explicit $\tau$ dependence in the functional determinants we can
write the relevant part of the $1$-loop correction  as
\begin{equation}
\tilde \Gamma_1=-\ln Z_1=-\frac{\T}{4 \pi}\int_{-\infty}^{\infty}d \omega\ \ln \frac{
\det^{8}[-\partial_1^2+\omega^2+\epsilon^2 \cos^2 \sigma] }{\ \det^{2}[-\partial_1^2+\omega^2+2 \epsilon^2
 \cos^2 \sigma]\ \det^{3}[-\partial_1^2+\omega^2]\ \det[Q_{\omega}]} \ , \label{abs}
\end{equation}
where $Q_{\omega}=Q(\partial_0 \rightarrow i \omega)$.

Let us now expand:
$Q_{\omega}= Q_{ \omega}^{(0)}+ \epsilon^2 Q_{ \omega}^{(2)} + ...\ , $
 where
\begin{eqnarray}
&&Q_{\omega}^{(0)}=\left(
               \begin{array}{ccc}
                 -(- \partial_1^2+\omega^2) & 0 & 0 \\
                 0 & -\partial_1^2+\omega^2-1 & -2 \omega \\
                 0 & 2 \omega & -\partial_1^2+\omega^2-1 \\
               \end{array}
             \right)\ , \la{aa} \\
&&Q_{ \omega}^{(2)}=\left(
               \begin{array}{ccc}
                 -\cos 2 \sigma & 0 & 2 \omega \sin \sigma \\
                 0 & \cos 2 \sigma+\frac{1}{2} & -{\omega} (\ha + \sin^2 \sigma) \\
                 -2 \omega \sin \sigma & {\omega} (\ha +\sin^2 \sigma) & \cos 2 \sigma-\frac{1}{2} \\
               \end{array}
             \right)\ .\la{bb}
\end{eqnarray}
Defining
\begin{eqnarray}
P_{\omega}=\left(
             \begin{array}{ccc}
               -(-\partial_1^2 + \omega^2)  & 0 & 0 \\
               0 & -\partial_1^2+\omega^2 & 0 \\
               0 & 0 & -\partial_1^2+\omega^2 \\
             \end{array}
           \right) \ , \la{puy}
\end{eqnarray}
the remaining  part  of $\tilde \G_1$ may be written as
\begin{eqnarray}
\frac{\T}{4 \pi}
\int d \omega \
\bigg(  \ln \frac{\det[Q_{\omega}]}{\det[Q_{\omega}^{(0)}]}-
\ln \frac{\det[ P_{\omega}]}{\det[Q_{\omega}^{(0)}]}\bigg) \ . \label{avd}
\end{eqnarray}
The second term  here  vanishes  for the same reason  why the rotation in \rf{rot}
lead to the standard massless kinetic terms for the two originally coupled  modes
and thus to the trivial   flat-space  partition function.  Indeed,
the ``mixed''  2 by 2  block contribution to
 $\ln \det[Q_{\omega}^{(0)}]$ can be
  written  as $\ln \det[-\partial_1^2+(\omega+i)^2] + \ln  \det[-\partial_1^2+(\omega-i)^2]$.
  Under the integral over $\omega$
   one can then shift $\omega$ by $- i$ in one term and by $+i$  in another
   to get the cancellation against other  massless determinants.
   These separate shifts are thus  consistent with the trivial (supersymmetric)
   result for $\G_1$ in flat space, and we shall perform  similar shifts of
   the corresponding terms   in what follows (in particular in  $\det[Q_{\omega}^{(0)}]$
   contribution of the first term in  \rf{avd}).

To compute the  first  term in (\ref{avd}) we expand in $\eps$ as in \rf{ll}
\begin{eqnarray}\la{ki}
\ln \frac{\det[Q_{\omega}]}{\det[Q_{\omega}^{(0)}]}=\epsilon^2 \Tr[(Q_{\omega}^{(0)})^{-1} Q_{\omega}^{(2)}]
+ ... =\epsilon^2 \sum_{n}\int_0^{2 \pi}\frac{d \sigma}{2 \pi}(Q_{\omega}^{(0)})^{-1}_{ij}
(Q_{\omega}^{(2)})_{ji} + ...\ .
\end{eqnarray}
The momentum-space propagator
 corresponding to $Q_{\omega}^{(0)}$  is
\begin{eqnarray}
(Q_{\omega}^{(0)})^{-1}=\left(
                           \begin{array}{ccc}
                             -\frac{1}{n^2+\omega^2} & 0 & 0 \\
                             0 & \frac{n^2+\omega^2-1}{n^4+2 n^2 (\omega^2-1)+(\omega^2+1)^2} & \frac{2 \omega}{n^4+2 n^2 (\omega^2-1)+(\omega^2+1)^2} \\
                             0 & -\frac{2 \omega}{n^4+2 n^2 (\omega^2-1)+(\omega^2+1)^2} & \frac{n^2+\omega^2-1}{n^4+2 n^2 (\omega^2-1)+(\omega^2+1)^2} \\
                           \end{array}
                         \right) \ .
\end{eqnarray}
It can be diagonalized  by a rotation
\begin{eqnarray}
M^{-1}(Q_{\omega}^{(0)})^{-1} M\equiv D^{(0)}_{\omega}=\left(
               \begin{array}{ccc}
                 -\frac{1}{n^2+\omega^2} & 0 & 0 \\
                 0 & \frac{1}{n^2+(\omega+i)^2} & 0 \\
                 0 & 0 & \frac{1}{n^2+(\omega-i)^2} \\
               \end{array}
             \right)    , \ \ \ \ \
M=\left(
    \begin{array}{ccc}
      1 & 0 & 0 \\
      0 & \frac{i}{2} & -\frac{i}{2} \\
      0 & \frac{1}{2} & \frac{1}{2} \\
    \end{array}
  \right) \la{my}
\end{eqnarray}
$Q_{\omega}^{(2)} $ gets rotated into
\begin{eqnarray}
M^{-1}Q_{\omega}^{(2)} M\equiv  D^{(2)}_{\omega}=\bp
                 -\cos 2 \sigma & \omega \sin \sigma & \omega \sin \sigma \\
                 -2 \omega \sin \sigma & {i \omega}(1- \ha\cos 2 \sigma)+\cos 2 \sigma & -\frac{1}{2} \\
                 -2 \omega \sin \sigma & -\frac{1}{2} & -{i \omega}(1-\ha\cos 2 \sigma)+\cos 2 \sigma
               \ep\no
\end{eqnarray}
and the $\eps^2$ term in  \rf{ki} becomes
\begin{eqnarray}
\epsilon^2 \sum_{n} \int_0^{2 \pi}\frac{d \sigma}{2 \pi} D^{(0)}_{\omega} D^{(2)}_{\omega}
=\epsilon^2 \sum_{n} \bigg[\frac{i \omega}
{n^2+(\omega+i)^2}-\frac{i \omega}{n^2+(\omega-i)^2}\bigg]\ . \label{qvw}
\end{eqnarray}
Thus finally  (using that $\k= \eps +..$, see \rf{abe})
\begin{equation}
\td E_1 = {\td \Gamma_1\ov \k \T} =-\frac{ \epsilon }{4 \pi}\int_{-\infty}^{\infty}
d \omega \sum_n
\bigg[\frac{2}{n^2+w^2}-\frac{i \omega}{n^2+(\omega+i)^2}+
\frac{i \omega}{n^2+(\omega-i)^2}\bigg] + O(\eps^3) \ . \label{qge}
\end{equation}
Doing the opposite shifts of $\omega$ in each of the last   two terms
we conclude that the order $\eps= \sqrt{2 \S} + ... $ term  in $\tilde E_1$ indeed vanishes, i.e.
\be  \tilde E_1 = 0 + O(\eps^3) \ . \la{eeq}   \ee

The  formal argument leading to \rf{eeq}
overlooked an  important subtlety of IR  divergences
that  we have so far postponed to  discuss but which will become  crucial
below.
 Indeed, if the sum over $n$ in \rf{qvw} runs over all  values
from $-\infty$ to  $+\infty$ one may get different results by interchanging the order
of integration over $\omega$ and summation over $n$: the integral over $\omega$
has an IR divergence at $n=0$.

In fact, as in  the usual  perturbative expansion
 near a soliton, there is an issue of possible IR   singularities due to
 a zero mode associated to the  translational symmetry $\s \to \s + \s_0$.
In the present case  of expansion in $\eps$  the ``free'' propagator
is essentially the massless
one on $R \times S^1$  and thus the zero mode
that is not damped in the path integral  corresponds to  $n=0$.
Its contribution can be either  regularized  by introducing a
small mass  or $i \epsilon$
in the propagator as in \ci{ftah}
 or by  isolating the  modes constant in $\s$
 in the path integral and thus
not including the $n=0$
 contributions in the propagators
 (as is  done, e.g., in quantizing a sigma model on a compact 2d
space).
This is  the prescription we shall
adopt  here, i.e.
 the sums over $n$  in \rf{mfg},\rf{mfg1},\rf{qvw}  and \rf{qge}
will be  understood  not to include the $n=0$ term.

\

Let  us now   include the  $\eps^2$   contribution to the effective action coming
from the $\r''$ term in the fermionic mass that we so far ignored, i.e.
compute $\delta E_1$  giving
\be  E_1= {\Gamma_1\ov \k \T}=  \tilde E_1 + \delta E_1  + O(\eps^3) \ , \ \ \ \ \ \  \delta E_1= O(\eps^2)  \ ,
   \la{oeq}   \ee
where (to the leading order of expansion of masses in $\eps$)
\begin{eqnarray}
\Gamma_1=-\frac{\T}{4 \pi}\int_{-\infty}^{\infty}d \omega\ \ln \frac{
\det^{4}[-\partial_1^2+\omega^2+\epsilon^2 \cos^2 \sigma -
 \epsilon \sin \s ] \det^{4}[-\partial_1^2+\omega^2+\epsilon^2 \cos^2 \sigma + \epsilon \sin \s
 ]}{\ \det^{2}[-\partial_1^2+\omega^2+2 \epsilon^2
 \cos^2 \sigma]\ \det^{3}[-\partial_1^2+\omega^2]\ \det[Q_{\omega}]} \nonumber
\end{eqnarray}
At order $\epsilon^2$  the $\r''= -\eps \sin \s  + ...$ part of the fermionic mass
contributes to $\G_1$  the following term
\bea
\delta  \Gamma_1 &=&
\frac{\T}{\pi} \epsilon^2 \int d \omega \sum_{n_1,n_2}
\frac{1}{n_1^2+ \omega^2} \frac{1}{n_2^2+\omega^2}\int \frac{d \sigma_1}{2 \pi}
\frac{d \sigma_2}{2 \pi}
\ \sin \sigma_1 \  \sin \sigma_2 \  e^{i (n_1-n_2) (\sigma_1-\sigma_2)} \no
\\
&=& \frac{\T}{\pi} \epsilon^2 \int d \omega \bigg[\sum_{n \neq 0, 1}
 \frac{1}{n^2 +\omega^2}\frac{1}{(n-1)^2 +\omega^2}+
\sum_{n \neq 0, -1} \frac{1}{n^2 +\omega^2}\frac{1}{(n+1)^2 +\omega^2}\bigg]\label{lk}
\la{gaga} \eea
Summing over $n$ gives
\begin{equation}
\delta  \Gamma_1 = - \frac{\T \epsilon^2}{4 \pi} \int
  d \omega \bigg[\frac{4}{\omega^2 (\omega^2+1)}- \frac{4 \pi \coth \pi \omega}{\omega (4 \omega^2+1)}\bigg] \ ,
\end{equation}
and thus  finally
\begin{equation}
\delta E_1= - \frac{ \epsilon}{4 \pi} \int d \omega \bigg[\frac{4}{\omega^2 (\omega^2+1)}-
\frac{4 \pi \coth \pi \omega}{\omega (4 \omega^2+1)}\bigg]= \epsilon (3 -4 \ln 2)
\ ,
\end{equation}
which leads to the value of $a_{01}$ quoted  in \rf{on}.\foot{As one can check, the same 
value is found by doing first the integral  and then the sum in \rf{gaga}
(this also applies to \rf{eeq},\rf{qge}). We thank M. Beccaria  for this observation.}

\renewcommand{\theequation}{4.\arabic{equation}}
 \setcounter{equation}{0}

\setcounter{equation}{0} \setcounter{footnote}{0}

\section{$1$- loop correction  to  the $S^{{3}/{2}}$ term in the string energy}

Let us  now compute the next  1-loop correction to the short string energy: 
the coefficient $a_{11}$
of the  $S^{{3}/{2}}$ term
in \rf{lmt} or  \rf{mt}.
For that we shall consider the next order of the
near flat space or $\eps \to 0$ expansion  of the fluctuation Lagrangian \rf{lag},\rf{feq}.
As in the previous section,  we  shall 
 treat separately the contributions coming  from the  $\rho''$ terms in the effective fermionic
 mass terms in \rf{sep}  (the  reason is that 
 while the expansion  of $\r'^2$  and similar mass terms contains 
 only even powers of $\epsilon$, the expansion of $\r''$  contains both even and odd powers 
 of $\epsilon$). 
 
 Let us first  ignore the contributions  coming from the 
 $\rho''$ terms   and add them later. 
As in (\ref{ll}) we shall use that
\begin{equation}
\ln \frac{\det [A+ \epsilon^2 B+\epsilon^4 C]}{\det A}=\epsilon^2 \Tr[ A^{-1} B]
-\frac{\epsilon^4}{2}\Tr[A^{-1}B A^{-1}B]+\epsilon^4 \Tr [A^{-1} C]+O(\epsilon^6) \label{abgg}
\end{equation}
 Expanding the fluctuation Lagrangian in $\eps$ using \rf{abe}, etc., we get
\begin{equation}
\tilde{L}=\tilde{L}_0+ \epsilon^2 \tilde{L}_1 + \epsilon^4 \tilde{L}_2 + ...\ ,
\end{equation}
where the $\epsilon^4$ terms in the masses and the mixing terms are
\bea
&&\delta \mu^2_t=\epsilon^4 (\frac{1}{2}-\cos^4 \sigma), \qquad
\delta\mu^2_{\phi}=\epsilon^4 (\frac{5}{32}-\cos^4 \sigma), \qquad \delta\mu^2_{\rho}=\epsilon^4
(\frac{21}{32}-\cos^4 \sigma) \ ,\no \\
&&
\delta\mu^2_{\beta} = - \epsilon^4 \cos^4 \sigma, \qquad \delta
\m^2_{\F}= -\frac{1}{2}\epsilon^4 \cos^4 \sigma \ , \\
&& \delta [4 \td {\rho} (\kappa \sinh \rho \partial_0 \td {t} -
w \cosh \rho \partial_0 \td {\phi})]=
-\frac{1}{2}\epsilon^4 \td {\rho} [(3+\cos 2 \sigma)\sin \sigma \partial_0 \td {t}
-(1 - {1\ov 8} \cos 4 \sigma)\partial_0 \td {\phi}] \no
\eea
Let us first compute the $\epsilon^4$
contribution  to 1-loop effective action
coming from the terms like $\epsilon^4 \Tr [A^{-1} C]$ in \rf{abgg}.
Using the  Fourier representation  in the (Euclidean) world-sheet time direction
($\partial_0 \rightarrow i \omega$)   the operator $Q$ acting on the $\td t, \td \r, \td \p$
subspace  can be
expanded  as (cf. \rf{aa},\rf{bb})
\bea \la{quu}
Q_{\omega}&=&Q^{(0)}_{\omega}+\epsilon^2 Q^{(2)}_{\omega}+\epsilon^4 Q^{(4)}_{\omega}+...\ , \\
Q^{(4)}_{\omega}&=&\bp
                 -(\frac{1}{2}-\cos^4 \sigma) & 0 & -\frac{\omega}{4}(3+ \cos 2 \sigma) \sin \sigma \\
                 0 & \frac{5}{32}-\cos^4 \sigma & -\frac{\omega}{32}(\cos 4 \sigma-8) \\
                 \frac{\omega}{4}(3+ \cos 2 \sigma) \sin \sigma & \frac{\omega}{32}(\cos 4 \sigma-8)
		   & \frac{21}{32}-\cos^4 \sigma \\
               \epm
	        \  . \la{pio}
\end{eqnarray}
As in \rf{my}
 we  rotate this to $M^{-1} Q^{(4)}_{\omega}M=D^{(4)}_{\omega}$ whose diagonal elements are
\begin{equation}
{\rm diag}[D^{(4)}_{\omega}]=
\Big\{
-\frac{1}{2}+\cos^4 \sigma; \frac{1}{32}(13 -8 i \omega-32 \cos^4 \sigma+i \omega \cos 4 \sigma);
\frac{1}{32}(13 +8 i \omega-32 \cos^4 \sigma-i \omega \cos 4 \sigma)\Big\} \no
\end{equation}
The computation of the $\eps^4$
term in (\ref{abgg}) coming from  the coupled bosonic part gives
\begin{eqnarray}
&&\Tr [(Q^{(0)}_{\omega})^{-1} Q^{(4)}_{\omega}]=
  \sum_n \int_0^{2 \pi} \frac{d \sigma}{2 \pi}\Tr [(Q^{(0)}_{\omega})^{-1} Q^{(4)}_{\omega}]=
   \sum_n \int_0^{2 \pi} \frac{d \sigma}{2 \pi} \Tr[D^{(0)}_{\omega} D^{(4)}_{\omega}] \nonumber\\
&=& \frac{1}{32} \sum_n \bigg[\frac{4}{n^2+w^2}+  \frac{1-8 i \omega}
{n^2+(\omega+i)^2}+\frac{1+8 i \omega}{n^2+(\omega-i)^2}\bigg]  \ . \label{nup}
\end{eqnarray}
The $\epsilon^4$ contribution
of the decoupled modes
 $\beta_u$  coming
  from the single insertion  of the $\epsilon^4$ perturbation,
  i.e. an $\epsilon^4 \Tr [A^{-1}C]$ type term is
\begin{eqnarray}
\frac{\det[ -\partial_1^2 +\omega^2+ 2 \epsilon^2 \cos^2
\sigma-\epsilon^4 \cos^4 \sigma]}{\det[-\partial_1^2 +\omega^2 ]}
\rightarrow &-&\epsilon^4  \sum_n \frac{1}{n^2 + \omega^2}\int_0^{2 \pi}
 \frac{d \sigma}{2 \pi} \cos^4 \sigma\nonumber\\
&=& -\epsilon^4  \frac{3}{8} \sum_n \frac{1}{n^2 + \omega^2} \label{wqr1} \ .
\end{eqnarray}
The single fermionic field gives   just half of this  contribution (up to the sign).

 Putting together all of the  contributions of the type $\epsilon^4 \Tr [A^{-1}C]$ we get
\begin{equation}
\tilde{\Gamma}_1 \rightarrow - \frac{\T \epsilon^4}{ 4  \pi}\int_{-\infty}^{\infty} d \omega \sum_{n}\bigg[
- { 7 \ov 8} \frac{28}{n^2+w^2} - { 1 \ov 32}    \frac{1-8 i \omega}{n^2+(\omega+i)^2}   - { 1 \ov 32}
\frac{1+8 i \omega}{n^2+(\omega-i)^2}\bigg] \ .   \label{qew}
\end{equation}
Now  let us
 compute the contributions of the type $\frac{1}{2} \epsilon^4\Tr[A^{-1}B A^{-1}B]$
 in \rf{abgg}.
Let us start with the decoupled fields ${\beta}_u$.  Using the form of
the $O(\epsilon^4)$ correction to the corresponding mass we get
\begin{eqnarray}
\Big(\frac{\epsilon^4}{2}Tr[A^{-1}B A^{-1}B]\Big)_\beta
&=& \frac{\epsilon^4}{2} \sum_{n_{1},n_{2}}\frac{1}{n_1^2+ \omega^2}\frac{1}{n_2^2+\omega^2}\no \\
&\times& 4\int_0^{2 \pi}\frac{d \sigma_1}{2 \pi} \frac{d \sigma_2}{2 \pi}
 \cos^2 \sigma_1\ e^{i \sigma_1 (n_1-n_2)} \cos^2 \sigma_2\ e^{-i \sigma_2 (n_1-n_2)}\la{qewa}
 \\
&=& \frac{\epsilon^4}{2}\sum_{n} \frac{1}{n^2+\omega^2}\bigg[\frac{1}{n^2+\omega^2}+
\frac{1}{4 [(n-2)^2+\omega^2]}+\frac{1}{4 [(n+2)^2+\omega^2]}\bigg]\nonumber
\end{eqnarray}
As discussed at the end of the previous section,
to project out the zero mode contribution  the sums  over $n$ in the massless propagators
  should not
include the $n=0$ point. Thus the sum in \rf{qew} should be over all $n\not=0$.
 In computing  the  integrals over $\s$ in \rf{qewa}    we have formally shifted
$n$  by $\pm 2$, so the last line  in the above  equation  should be understood as
a combination of the three sums where in the first sum  $n\not=0$, in the second
 $n\not=0,2$ and in the third  $n\not=0,-2$.

The corresponding fermionic contribution is essentially
 $\frac{1}{4}$ of \rf{qew}, as $\m_\F^2$ (without $\pm \r''$ part) 
  is half of $\mu^2_{\beta}$,
  but here there are two mass insertions.
 Putting together such contributions from the
   decoupled bosons and the fermions
 we observe that they  cancel each other.

Next,  let us find
the   $ \epsilon^4\Tr[A^{-1}B A^{-1}B]$ type   contribution of
 the coupled set of fluctuations.
 It   can be   written as (see \rf{aa},\rf{bb})
\begin{eqnarray}
&&\frac{\epsilon^4}{2}\Tr[(Q^{(0)}_{\omega})^{-1}Q^{(2)}_{\omega} (Q^{(0)}_{\omega})^{-1}
Q^{(2)}_{\omega}]\la{tak}\\
&=&\frac{\epsilon^4}{2} \sum_{n_{1},n_{2}}\int_0^{2 \pi} \frac{d \sigma_1}{2 \pi}\frac{d \sigma_2}{2 \pi}
 \Tr[(Q^{(0)}_{\omega})^{-1}(n_1)Q^{(2)}_{\omega}(\sigma_2) (Q^{(0)}_{\omega})^{-1}(n_2)
 Q^{(2)}_{\omega}(\sigma_1)]\ e ^{i (n_1-n_2) (\sigma_1-\sigma_2)} \no
\end{eqnarray}
To compute this expression  we again  first  diagonalize the propagator matrix
and then integrate over $\s$. Putting together all the contributions from the two
insertions of the $\eps^2$ perturbations
  and adding the contribution with single $\eps^4$
  insertion (\ref{qew}) we get  the following result
  for the $1$-loop effective action to order $\epsilon^4$
  (without yet including the $\r''$  fermionic mass term contributions)
\begin{eqnarray}
\tilde{\Gamma}_1{(\eps^4)} &=& -\frac{\T \epsilon^4}{4 \pi}\int_{-\infty}^{\infty} d
 \omega \bigg\{\sum_{n}\bigg[-\frac{7}{8}\frac{1}{n^2+w^2}-\frac{1}{32}
 \frac{1-8 i \omega}{n^2+(\omega+i)^2}-\frac{1}{32}\frac{1+8 i \omega}{n^2+(\omega-i)^2}\bigg]\nonumber\\
&+&\frac{1}{2}\sum_n \bigg[-\frac{ \omega^2}{[n^2+(\omega+i)^2]^2}-
\frac{ \omega^2}{[n^2+(\omega-i)^2]^2}\nonumber\\
&+& \frac{1}{4} \frac{1}{n^2+w^2}\bigg(\frac{1}{(n-2)^2+\omega^2}+\frac{1}
{(n+2)^2+\omega^2}\bigg)+\frac{1}{2}\frac{1}{[n^2+(\omega+i)^2][n^2+(\omega-i)^2]}\nonumber\\
&+& \omega^2\bigg(\frac{1}{(n+1)^2 +\omega^2}+\frac{1}{(n-1)^2+\omega^2}\bigg)
\bigg(\frac{1}{n^2+(\omega+i)^2}+\frac{1}{n^2+(\omega-i)^2}\bigg)\nonumber\\
&+& \frac{(1+\frac{i \omega}{2})^2}{4} \frac{1}{n^2+(\omega-i)^2}\bigg(\frac{1}
{(n-2)^2+(\omega-i)^2}+\frac{1}{(n+2)^2+(\omega-i)^2}\bigg)\nonumber\\
&+&\frac{(1-\frac{i \omega}{2})^2}{4} \frac{1}{n^2+(\omega+i)^2}\bigg(\frac{1}
{(n-2)^2+(\omega+i)^2}+\frac{1}{(n+2)^2+(\omega+i)^2}\bigg)\bigg]\bigg\}\label{qio}
\end{eqnarray}
Again, this expression   should be understood as a combination
of  sums over $n$ where  the values of $n$ for which the effective (shifted) value of $n$
vanishes  should be projected out as it came  from
 the original $n_i$   in the propagator after doing the
integral over $\s$ and shifting the  summation index.  For example, we have
\begin{eqnarray}
&& \sum_{n_{1}\neq 0 ,n_{2}\neq 0} \frac{1}{n^2_1+\omega^2} \frac{1}{n_2^2+
 \omega^2}\int_0^{2 \pi} \frac{d \sigma_1}{2 \pi} \frac{d \sigma_2}{2 \pi} \cos 2 \sigma_1 \cos 2 \sigma_2
 \  e^{i (n_1 - n_2) (\sigma_1-\sigma_2)}\nonumber\\
&=&\frac{1}{4}\sum_{n \neq 0, 2} \frac{1}{n^2+\omega^2} \frac{1}{(n-2)^2+\omega^2}+
\frac{1}{4}\sum_{n \neq 0,  -2} \frac{1}{n^2+\omega^2} \frac{1}{(n+2)^2+\omega^2} \ . \la{jjj}
\end{eqnarray}
The first three terms in \rf{qio}  can be   simplified  as in \rf{qge}
 by doing  separate shifts of $w$ by $\pm i$ in the last two  terms; this gives
\begin{equation}
-\frac{1}{32}\sum_{n\not=0}  \int_{-\infty}^{\infty} d \omega \bigg[\frac{28}{n^2+w^2}+\frac{1-
8 i \omega}{n^2+(\omega+i)^2}  + \frac{1+8 i \omega}{n^2+(\omega-i)^2}\bigg]
= - \frac{7}{16}\sum_{n\neq 0} \int_{-\infty}^{\infty} d \omega \frac{1}{n^2+ \omega^2}.
\end{equation}
Similar separate shifts  of $w$ under the
integral $\int^\infty_{-\infty} d \omega$ can
 be used to transform  some other terms  in \rf{qio}.
 For example, we get
\begin{equation}
 \frac{ \omega^2}{[n^2+(\omega+i)^2]^2}+\frac{ \omega^2}{[n^2+(\omega-i)^2]^2}\rightarrow
  2  \frac{\omega^2-1}{(n^2+\omega^2)^2} \label{qar} \ ,
\end{equation}
\begin{equation}
 \frac{1}{[n^2+(\omega+i)^2][n^2+(\omega-i)^2]}
 = { i \ov 2 \omega} \Big[ { 1 \ov n^2+(\omega+i)^2} - { 1 \ov n^2+(\omega-i)^2}\Big]
  \rightarrow
  -\frac{1}{2(\omega^2+1)}\frac{1}{n^2 +  \omega^2} \no
\end{equation}
Using  the identity  $\frac{1}{a  b}=(\frac{1}{a}-\frac{1}{b})\frac{1}{b-a}$
with $a,b$ being $(n+k)^2 + (\omega+ v)^2$,\  ($k=0,\pm 2$,\  $v=0, \pm i$)
and shifting $\omega$ in terms  containing
 only propagator  factors  with  $(\omega \pm i)$
 one finds that
\begin{eqnarray}
&&\sum_{n\not=0,-1} \frac{\omega^2}{[(n+1)^2 +\omega^2]
[n^2+ (\omega+i)^2]} + \sum_{n\not=0,1}
\frac{\omega^2}{[(n-1)^2 +\omega^2] [n^2+(\omega+i)^2]} + c.c.
    \nonumber\\
&& \rightarrow
\sum_{n\neq 0,1} \frac{\omega^2 (n-1)}{[(n-1)^2+\omega^2][(n-1)^2+\omega^2]}-\sum_{n\neq 0,-1} \frac{\omega^2 (n+1)}{[(n+1)^2+\omega^2]
[(n+1)^2+\omega^2]}\nonumber\\
&& \ \ \ \ \ \ \ \ \ \ -\sum_{n\neq 0,-1}\frac{n-\omega^2 (n+2)}{(n^2+\omega^2)^2}-\sum_{n\neq 0,1} \frac{-n+ \omega^2 (n-2)}{(n^2+\omega^2)^2}\nonumber\\
&&\ \ \ \ \ \ = -\frac{2}{(\omega^2+1)^2}+ \sum_{n\neq 0} \frac{4 \omega^2}{(n^2+\omega^2)^2}  \la{nad} \ .
\end{eqnarray}
The second line above comes from the unshifted terms, while the
 third line from the $\omega$-shifted terms. Let us mention that to arrive
 to the result in the last line in (\ref{nad}) we have assumed the prescription in which
  the sums over $n$  (in infinite limits)  are computed {\it before}
  doing the  integral over $\omega$ so that one is allowed to do shifts of the summation index $n$.
   If one would  instead assume that  the integral over  $\omega$ (in infinite limits)
    is done  before the evaluation of the sums  the
    result would be  different.\footnote{We are grateful to
  M. Beccaria for pointing out  this ambiguity to us.}
  We shall discuss the  origin of  this ambiguity in Appendix C.

Performing similar shifts of $\omega$ and  $n$ in  the last two  lines in \rf{qio} we get
\begin{eqnarray}
&& \frac{(1+\frac{i \omega}{2})^2}{4}
\frac{1}{n^2+(\omega-i)^2}\bigg[
\frac{1}{(n-2)^2+(\omega-i)^2}+\frac{1}{(n+2)^2+(\omega-i)^2}
\bigg] + c.c. \nonumber\\
&\rightarrow& -
 \frac{\omega^2-1}{4(n^2+\omega^2)((n-2)^2+
\omega^2)} \ ,  \la{ku}
\end{eqnarray}
where the final term should be summed over $n\neq 0,2$.

Collecting the above expressions we get for \rf{qio}
\begin{equation}
\tilde{\Gamma}_1(\eps^4)  = -\frac{\T \epsilon^4}{4 \pi}\int_{-\infty}^{\infty} d \omega\   \Big(C_0 + C_1 +
 C_2 + \sum_{n=3}^{\infty} S_n\Big) \ ,  \label{rem}
\end{equation}
where
\begin{equation}
C_0=-\frac{1}{(\omega^2+1)^2} \ ,\ \
 \ \ \ C_1= \frac{7 \omega^4 +84 \omega^2
  +93}{8 (\omega^2+1)^2 (\omega^2+9)} \ , \ \ \ \ \ C_2= \frac{8 \omega^6 +137 \omega^4 -89 \omega^2 -308}{8 (\omega^2+4)^2 (\omega^4 +17 \omega^2+16)}
\end{equation}
\bea
&&S_n=\frac{1}{8 (n^2+\omega^2)^2}\bigg[9 \omega^2 -7 n^2 +16 -
\frac{2 (n^2+\omega^2)}{\omega^2+1} \no \\
&& \ \ \ \ \  \ \ \ \ \ \ \
 \ \ \ \ \  \-\ (n^2+\omega^2)(\omega^2-3)\bigg(\frac{1}{(n+2)^2+\omega^2}+
\frac{1}{(n-2)^2+\omega^2}\bigg)\bigg] \ . \la{fh}
\eea
The result is  UV  finite as expected  \ci{ft1}. It is also
IR finite  (which would not be the case if the zero mode contributions
were not properly projected out).

The integrals over $\omega$ give
\bea
&&\int_{-\infty}^{\infty} d \omega\ C_0= -\frac{\pi}{2},\ \ \ \
\ \ \int_{-\infty}^{\infty} d \omega\ C_1= \frac{9 \pi}{8}, \ \ \
\quad \int_{-\infty}^{\infty} d
\omega\ C_2= \frac{17 \pi}{128}
\  .
\eea
As discussed above we need first to compute the sum and then the integral.
Remarkably, the sum over $n$ of $S_n$ in  \rf{fh} can be performed exactly
and we obtain
\begin{eqnarray}
C_3(\omega)& \equiv & \sum_{n=3}^{\infty}S_n = \frac{\pi^2 (\omega^2+1) \mbox{csch}^2 \pi \omega}{2 \omega^2}+\frac{\pi (5 \omega^2 +4) \coth \pi \omega}{8 \omega^3 (\omega^2+1)}-\frac{53}{48 (\omega^2+1)}-\frac{27}{32 (\omega^2+4)}\nonumber\\
&+&\frac{3}{16 (\omega^2+9)}+\frac{19}{96 (\omega^2+16)}-\frac{5}{8 \omega^2}-\frac{1}{4 (\omega^2+1)^2}+\frac{6}{(\omega^2+4)^2}-\frac{1}{\omega^4} \ .
\end{eqnarray}
To compute $\int_{-\infty}^{\infty} d  \omega C_3(\omega)$ it is convenient to decompose\foot{We thank
M. Beccaria  for this observation and correcting our original result for $a_{11}$.} $C_3 = C_{30}+ C_{31}$ as
\begin{eqnarray}
C_{30}(\omega)& = & \frac{\pi^2  \mbox{csch}^2 \pi \omega}{2 \omega^2}+\frac{\pi (5 \omega^2 +4) \coth \pi \omega}{8 \omega^3 (\omega^2+1)}-\frac{53}{48 (\omega^2+1)}-\frac{27}{32 (\omega^2+4)}\nonumber\\
&+&\frac{3}{16 (\omega^2+9)}+\frac{19}{96 (\omega^2+16)}-\frac{1}{8 \omega^2}-\frac{1}{4 (\omega^2+1)^2}+\frac{6}{(\omega^2+4)^2}-\frac{1}{\omega^4} \ ,
\end{eqnarray}
\begin{equation}
C_{31}(\omega)=\frac{\pi^2}{2}  \mbox{csch}^2 \pi \omega - \frac{1}{2 \omega^2} \ .
\end{equation}
The first integral can be performed using residues theorem on a contour
 that includes the real axis and a semi-circular loop going to infinity in 
 the upper half plane; the simple poles are at $\omega= i n$, $n >0$. We obtain
\begin{equation}
\int_{-\infty}^{\infty} d \omega \ C_{30}= -\frac{133}{128}\pi + \pi \zeta(3)\ .
\end{equation}
Noticing that $C_{31}=\frac{d}{d \omega}(\frac{1}{2\omega}-\frac{\pi}{2}\coth 
\pi \omega)$ the other integral is just
\begin{equation}
\int_{-\infty}^{\infty} d  \omega \ C_{31}=-\pi\ .
\end{equation}
Collecting these expressions  we obtain the following result for \rf{qio}
\be
\tilde{\Gamma}_1(\epsilon^4) 
=  { \T \ov 4} \big[ {41 \ov 32} - \zeta(3)   \big]\ \eps^4 \ .\la{eqr} \ee
\bigskip

Let us now include 
 the extra contributions due to   the $\rho''$ terms  in the fermionic masses
 in \rf{sep}: according to \rf{fero} the non-trivial part of the 
  full   1-loop fermionic contribution is
 (after Fourier transform in time direction)
\begin{equation}
- \ha \ln \Big( \frac{\det^4 [-\partial_1^2 + \omega^2 + \rho'^2 + \rho'']}
{\det^4 [-\partial_1^2 +\omega^2]}\ 
\frac{\det^4 [-\partial_1^2 + \omega^2 + \rho'^2 - \rho'']}{\det^4 [-\partial_1^2 +\omega^2]} 
\Big) \ .
 \label{eio}
\end{equation}
Recalling that 
\begin{equation}
\rho'^2= \epsilon^2 \cos^2 \sigma - \frac{\epsilon^4}{2}\cos^4 \sigma + ... , 
\quad \quad  \rho''=-\epsilon \sin \sigma + \frac{3 \epsilon^3}{4}\sin \sigma \cos^2 \sigma + ...\
, 
\end{equation}
and since the full expression is symmetric under $\r'' \to - \r''$ 
we conclude  that $\rho''$ terms contribute only at even orders in $\epsilon$. The extra contributions that we need to compute at order $\epsilon^4$ are the following
\begin{eqnarray}
&& \ln \frac{\det [A + B \epsilon + C \epsilon^2 + D \epsilon^3 + E \epsilon^4]}{\det A} \rightarrow \epsilon^4
Tr[(A^{-1}B)^2 A^{-1} C] - \epsilon^4 Tr[A^{-1} B A^{-1} D]\nonumber\\
&-& \frac{\epsilon^4}{4}Tr[(A^{-1} B)^4]
=\epsilon^4( \rI + \II+ \III)\ , 
\end{eqnarray}
where in our case we have
\begin{equation}
B=-\sin \sigma, \quad \quad C= \cos^2 \sigma, \quad \quad D=  \frac{3}{4}\sin \sigma \cos^3 \sigma, \quad \quad E= -\frac{1}{2}\cos^4 \sigma
\end{equation}
The additional contribution to (\ref{eio}) is then 
\begin{equation}\la{degam}
\delta \Gamma_1(\epsilon^4) = - \frac{\T \epsilon^4}{4 \pi}\int d \omega \ 8\  (\rI+\II+\II)\ . 
\end{equation}
The explicit computation of these terms  gives
\begin{eqnarray}
\rI &=& \frac{1}{8}\bigg[\sum_{n \neq 0,1} \frac{1}{(n^2+\omega^2)^2}\frac{1}{(n-1)^2 +\omega^2}+\sum_{n \neq 0,-1} \frac{1}{(n^2+\omega^2)^2}\frac{1}{(n+1)^2 +\omega^2}\bigg]\nonumber\\
& & \ - \frac{1}{16}\bigg[\sum_{n \neq 0,1,2} \frac{1}{n^2 +\omega^2}
\frac{1}{(n-1)^2+\omega^2}\frac{1}{(n-2)^2+\omega^2}\nonumber\\
& & \ +\sum_{n \neq 0,-1,-2}
 \frac{1}{n^2 +\omega^2}\frac{1}{(n+1)^2+\omega^2}\frac{1}{(n+2)^2+\omega^2}\bigg]\\
&=&\frac{\pi (12 \omega^4+23 \omega^2 +2)}{16 \omega^3 (4 \omega^2+1)^2
 (\omega^2+1)}\coth \pi \omega + \frac{\pi^2}{8 \omega^2 (4 \omega^2+1)}
 \frac{1}{\sinh^2 \pi \omega}-\frac{\omega^4 +12 \omega^2 +8}{8 \omega^4
  (\omega^2+1)^2(\omega^2+4)} \ , \nonumber\\
\II &=&\frac{3}{64}\bigg( \sum_{n \neq 0,1} \frac{1}{n^2+\omega^2}\frac{1}{(n-1)^2+\omega^2}+ \sum_{n \neq 0,-1} \frac{1}{n^2+\omega^2}\frac{1}{(n+1)^2+\omega^2}\bigg)\nonumber\\
&=&\frac{3 }{64}\bigg(\frac{4 \pi }{\omega (1+ 4 \omega^2)}\coth \pi \omega - \frac{4}{\omega^2 (1+ \omega^2)}\bigg) \ ,
\end{eqnarray}
and 
\begin{eqnarray}
\III &=&- \frac{1}{64}\bigg[\sum_{n \neq 0,-1,-2} \frac{1}{n^2+\omega^2}\frac{1}{((n+1)^2
+\omega^2)^2}\frac{1}{(n+2)^2+\omega^2}\no \\
&+&
\sum_{n \neq 0,1,2} \frac{1}{n^2+\omega^2}\frac{1}{((n-1)^2+\omega^2)^2}\frac{1}{(n-2)^2+\omega^2}\nonumber\\
&+& \sum_{n \neq 0,1} \frac{1}{(n^2+\omega^2)^2}\frac{1}{((n-1)^2+\omega^2)^2}+2 \sum_{n \neq 0,1,-1} \frac{1}{(n^2+\omega^2)^2}\frac{1}{(n-1)^2+\omega^2}\frac{1}{(n+1)^2+\omega^2}\nonumber\\
&+&\sum_{n \neq 0,-1} \frac{1}{(n^2+\omega^2)^2}\frac{1}{((n+1)^2+\omega^2)^2}\bigg]\nonumber\\
&=& - \frac{\pi^2}{32}\frac{8 \omega^4 + 10 \omega^2 + 2}{\omega^2 (4 \omega^2+1)^3 (\omega^2+1)}\frac{1}{\sinh^2 \pi \omega}-\frac{\pi}{32}\frac{60 \omega^4 +35 \omega^2 +2}{\omega^3 (4 \omega^2+1)^3 (\omega^2+1)}\coth \pi \omega \nonumber\\
&+& \frac{\omega^2 +2}{4 \omega^4 (\omega^2+1)^2 (\omega^2+4)}\  . 
\end{eqnarray}
According to the prescription  already used above 
 we have performed the sums first and then the integral over $\omega$.
Collecting the three contributions we obtain for \rf{degam}
\begin{eqnarray}
&&\delta \Gamma_1(\epsilon^4)= - \frac{\T \epsilon^4}{4 \pi}\int d \omega \ 
\bigg[\frac{\pi^2 (8 \omega^2+1)}{2 (4 \omega^3+\omega)^2}
\frac{1}{\sinh^2 \pi \omega} \no \\
&&+\frac{\pi (96 \omega^8+240 \omega^6+202 \omega^4 
+33 \omega^2+2)}{4 (\omega^2+1)(4 \omega^3+\omega)^3}\coth \pi \omega 
+ \frac{3 \omega^6 +17 \omega^4 +32 \omega^2 +8}{2 \omega^4 (\omega^2+1)^2 (\omega^2+4)}\bigg]
\end{eqnarray} 
Performing the integral over $\omega$ we find 
\begin{equation}\la{jp}
\delta \Gamma_1 (\epsilon^4)= \frac{\T}{4}\bigg[-\frac{559}{72}+ 6 \ln 2 + \frac{5}{2}\zeta(3)\bigg] \epsilon^4
\end{equation}

To obtain the total $1$-loop energy we need to
add the contributions  in \rf{eqr} and \rf{jp}, 
 $\tilde{\Gamma}_1(\epsilon^4)+ \delta \Gamma_1(\epsilon^4)$, and also to 
express  $\kappa$ and $\epsilon$ in terms of the spin.
 As a result, we  finally obtain the following 
 $1$-loop correction 
\begin{equation}
E_1 =\sqrt{2 \mathcal{S}} (3- 4 \ln 2)+  \frac{1}{\sqrt{2}}\bigg[\frac{-1219 + 864 \ln 2 + 432 \zeta(3)}{288}\bigg]\mathcal{S}^{3/2}+O(\mathcal{S}^{5/2}) \ ,
\end{equation} 
which leads to the value of the coefficient $a_{11}$  quoted in (\ref{on}).

\

\section*{Note Added }
A  different  calculation of the 1-loop correction to the folded string 
energy in the small spin limit was recently carried out, 
partly using numerical evaluation, by N. Gromov (private communication).
It led to the same  structure of the expansion of the energy \rf{lmt}  as found here 
but apparently with somewhat different coefficients than in \rf{on}. 
This disagreement may  be related to our prescription of projecting out the 
zero mode contribution.

\bigskip

\section*{Acknowledgments }

We thank   V. Forini,  N. Gromov, M. Kruczenski,  J. Plefka  and    R. Roiban  for
useful  discussions.
We are especially grateful to M. Beccaria for pointing  out an error in our original computation
and for many important discussions  and clarifications.
 A.T. was supported in part by NSF under grant PHY-0653357.
  A.A.T. acknowledges the support of  STFC.

\bigskip

\appendix
\subsection*{Appendix A:  Vanishing of $1$-loop correction to \\
 folded string
energy    in flat space
}
\refstepcounter{section}
\def\theequation{A.\arabic{equation}}
\setcounter{equation}{0}

Here  we shall show the  vanishing
of  the $1$-loop correction to the folded string energy  in flat space by using the GS formalism
in the covariant $\kappa$-symmetry gauge.

In the flat space
\begin{equation}
ds^2 = -dt^2 + d \rho^2 + \rho^2 d \phi^2
\end{equation}
the folded string solution in the conformal gauge is  (here  $\eps$ is an arbitrary constant)
\begin{equation}
\bar t =\eps  \tau, \quad \quad \bar \rho= \eps  \sin \sigma, \quad \quad \bar \phi=\tau \ .
\end{equation}
Equivalently, $x_1= \r \cos \phi =   \eps  \sin \sigma \ \cos \tau , \ \
x_2= \r \sin \phi=   \eps \sin \s \ \sin   \tau$.
The bosonic part of  quadratic fluctuation Lagrangian in the conformal gauge is
\begin{equation}
\tilde{L}_B= \dot{\tilde{t}}^2 - \tilde{t}'^2 - \dot{\tilde{\rho}}^2 + \tilde{\rho}'^2 - \tilde{\rho}^2 - \bar \r^2
 (\dot{\tilde{\varphi}}^2- \tilde{\varphi}'^2)
-4 \bar \r \tilde{\rho}\dot{\tilde{\phi}}
\end{equation}
After the  rescaling $\bar \r  \tilde{\phi}  =\tilde{\varphi}$ this becomes
\begin{equation}
\tilde{L}= \dot{\tilde{t}}^2 - \tilde{t}'^2- \dot{\tilde{\rho}}^2 + \tilde{\rho}'^2 - \tilde{\rho}^2-\dot{\tilde{\varphi}}^2+ \tilde{\varphi}'^2-\tilde{\varphi}^2 - 4 \tilde{\rho} \dot{\tilde{\varphi}}
\end{equation}
Performing further the rotation
$
\tilde{\rho}=\eta_1 \cos \tau + \eta_2 \sin \tau, \ \  \tilde{\varphi}=-\eta_1 \sin \tau + \eta_2 \cos \tau,
$
this  becomes the  Lagrangian for  free massless bosons
\begin{equation}
\tilde{L}_B= -\partial_a \tilde{t} \partial^a \tilde{t}  +  \partial_a \eta_1 \partial^a \eta_1 + \partial_a \eta_2 \partial^a \eta_2
\end{equation}
Starting with the  quadratic part of the GS superstring action in fermions in flat space in general coordinates
\begin{equation}
L_F= (\sqrt{-g}g^{ab}\delta^{IJ}- \epsilon^{ab}s^{IJ})\bar{\theta}^{I} \vr_a D_b \theta^{J}  \ ,
\end{equation}
where $s^{IJ}= {\rm diag} (1,-1)$, \  $\vr_a= \Gamma_{A} E_{\mu}^{A}\partial_a X^{\mu}$,
 $D_a = \partial_a + \frac{1}{4} \del_a X^M \omega_M^{AB} \Gamma_{AB}$,   and
  fixing  the  $\kappa$-symmetry gauge as
\begin{equation}
\theta_1=\theta_2 = \theta
\end{equation}
we get
\begin{equation}
L_F= 2i \sqrt{-g}g^{ab} \bar{\theta} \vr_a D_b \theta   \ .
\end{equation}
The induced metric for the above  classical solution is, of course,  conformaly flat
\begin{equation}
ds^2= \eps^2 \cos^2 \sigma (- d \tau^2 + d \sigma^2)  \ .
\end{equation}
Then labeling the coordinates as $(X^0, X^1, X^2) = (t, \r, \phi)$ we get
\bea
\vr_0 = \eps ( \Gamma_0 + \sin \sigma\ \Gamma_2) \ , \quad \quad \rho_1= \eps  \cos \sigma\ \Gamma_1 \, \no
\\
D_0 = \partial_0 - \frac{1}{2}\Gamma_{12}\ , \quad \quad D_1=\partial_1 \ .
\eea
and thus
\bea
&&L_F = 2 \bar{\theta} D_\F \theta \ ,  \\
&& \hat  D_\F= i (- \vr_0 D_0 + \vr_1 D_1)=
 i \eps \Big[ - ( \Gamma_0 +  \sin \sigma\  \Gamma_2)\partial_0 +  \Gamma_1 \cos \sigma\  \partial_1
 + \frac{1}{2}(\Gamma_{012}-\Gamma_1 \sin \sigma)\Big] \no
\eea
After the rotation
\begin{equation}
\tilde\theta = e^{-\frac{1}{2}\a \Gamma_0 \Gamma_2} {\theta}, \quad  \quad \sinh  \a = \tan \sigma \ ,
\end{equation}
the fermionic operator becomes
\begin{equation}
\td D_F= i \eps \Big( - \Gamma_0 \cos \sigma\  \partial_0 + \Gamma_1 \cos \sigma\ \partial_1 - \frac{1}{2}\Gamma_1 \sin \sigma
\Big) \  .
\end{equation}
Finally, rescaling  $\tilde{\theta}=\frac{1}{\sqrt{\eps \cos \sigma}}\vt$ we end up with
 the  action for $\vt$   with the free massless Dirac operator
\begin{equation}
D_\F= i (- \Gamma_0 \partial_0 +\Gamma_1 \partial_1)  \ .
\end{equation}
Thus both the bosonic and the fermionic fluctuations decouple from the background
and  cannot contribute to the classical relation between the energy and the spin, $E= \sqrt{ {2 \ov \a'} S}$.
In fact, as is well known,
 the 1-loop  shift of the GS superstring
 vacuum energy  iz zero because of the balance of the number of
bosonic and fermionic degrees of freedom (assuming we include also the conformal gauge ghost contribution).

\appendix
\subsection*{Appendix B:  Generalization
 to non-zero $S^5$ angular momentum
}
\refstepcounter{section}
\def\theequation{B.\arabic{equation}}
\setcounter{equation}{0}

The  above  discussion of the spinning string in $AdS_5$
 can be generalized to the case of the $(S,J)$ string
which is spinning with spin $S$  in $AdS_5$   but   also moving with momentum $J$
around big circle  in $S^5$ \ci{ft1}.
This generalization is potentially important as it allows one to relate the corresponding
string states to operators like $\tr (D_+^S \Phi^J)$
in the closed $sl(2)$  sector of the SYM theory  (with $J$ having the interpretation
of the  length of the corresponding spin chain \ci{bs}).

The relations in section 2 have straightforward generalization
to the case  when the string also moves along the $S^1 $ in $S^5$:
\bea
&&\varphi= \nu \tau, \ \ \ \ \   J= \sql\ \nu  \ , \  \ \ \ \ \ \ \
\rho'^2 = \kappa^2  \cosh^2 \rho - w^2 \sinh^2 \rho - \nu^2 \ , \\
&&0 \leq \r \leq \r_* \ , \ \   \ \ \  \ \
\coth^2 \rho_* = \frac{w^2-\nu^2}{\kappa^2-\nu^2}\equiv 1+ \frac{1}{\epsilon^2} \ ,
\ \ \ \ \r_*= \eps - { 1 \ov 6}{ \eps^3} + ... \ .
\eea
Here $\nu \equiv \J$ plays the role of the semiclassical $S^5$
momentum parameter and
 $\eps$ again measures the length of the string.
To include  nonzero
 $\nu$ one is  to  shift
 $ w\ \to\ \sqrt{ w^2 - \nu^2}, \  \ \k\ \to\ \sqrt{ \k^2 - \nu^2}$.
 We get (cf. \rf{uu},\rf{qdr})  \ci{ft1}
 \bea\la{pion}
&& \sqrt{\kappa^2-\nu^2}=\epsilon \
 _2 F_1(\frac{1}{2},\frac{1}{2};1;-\epsilon^2) \ , \  \ \ \  \
  \ \ \mathcal{E}_0=\frac{\kappa}{\sqrt{\kappa^2-\nu^2}}
\epsilon \
  _2F_1(-\frac{1}{2},\frac{1}{2};1;-\epsilon^2) , \\
&&
 \mathcal{S}=\frac{w}{\sqrt{\kappa^2-\nu^2}}\frac{\epsilon^3}{2}
 \ _2F_1(\frac{1}{2},\frac{3}{2};2;-\epsilon^2) \ .
\eea
To consider the short string limit we should expand  in small $\eps$ while keeping
 $\nu$  arbitrary.  Then  we find
 \be
 \mathcal{E}^2_0=
\nu^2 + \epsilon^2 (1+\nu^2) + \frac{\epsilon^4}{2}(1+\frac{\nu^2}{4})+O(\epsilon^6)  \ , \ \ \
\S^2 = \frac{\epsilon^4}{4} (1 + \nu^2) + \frac{\epsilon^6}{16} (1 - \nu^2) +
O(\epsilon^8) \ee
i.e.
\be
\epsilon^2 =\frac{2 \mathcal{S}}{\sqrt{1 +\nu^2}}  + O(\S^2) \ , \ \ \ \ \ \ \ \ \ \
 \mathcal{E}^2_0 = \nu^2 + 2 \mathcal{S}\sqrt{1+\nu^2}   + O(\S^2)  \ . \la{maq} \ee
The short string limit $\eps \ll 1$  \ci{ft1}
can   thus be achieved  by, e.g.,
  considering  a slowly spinning string $\S \ll 1$  or by
assuming large momentum
in $S^5$, i.e.  $\nu \gg 1$.  The latter is the fast string  or BMN-like
 limit  while the former may be   called
 a near flat space  limit in which $\nu$   may be  kept arbitrarily small.

Below we shall concentrate on  the short string limit $\epsilon\ll 1$. If we
 further assume that
 $\epsilon \ll \nu$
   then
 the classical energy  will be
\begin{equation}
\mathcal{E}_0= \nu + \frac{\S}{\nu}\sqrt{\nu^2+1}+O(\S^2) \ .  \label{aml}
\end{equation}
If we  then  expand in large $\nu \gg 1$ that will correspond to  the usual
fast short string limit  where one takes $\nu$ large at fixed
${\S\ov \nu}= {S\ov J}$ and then expands in ${
\S\ov \nu }\ll 1$  \ci{ft1}
 \be  \E_0= \nu + \S  + { \S \ov 2 \nu^2 } + ...  \ , \ \ \ \ \ \ \ \
    \nu \gg 1 , \  \ \ { \S \ov \nu } \ll 1 \ \label{jue} . \ee
In the slow short string limit we have $\eps \ll 1, \ \S \ll 1$;
 if we
assume in addition that the $S^5$
rotational
energy is smaller than the spinning  one,
then  $ \nu \ll \sqrt \S \ll 1$.
In this case   $\nu \ll \eps$  which is  opposite to the
above assumption that led to \rf{aml}. Here we get
$  \epsilon=
 \sqrt{2 \mathcal{S}}-
 \frac{1}{4 \sqrt{2}}\mathcal{S}^{{3}/{2}} \big(1 +\frac{2\nu^2}{\S}\big)
 + ...
 $ so that   the classical energy has
 a ``near flat space'' expansion form
 \begin{eqnarray}
\mathcal{E}_0=\sqrt{2 \mathcal{S}}\ \big(1+ \frac{\nu^2}{4\mathcal{S}}
+...\big)
+\frac{3}{4 \sqrt{2}}\mathcal{S}^{{3}/{2}}\big(1+ \frac{5\nu^2}{12\mathcal{S}}
+...\big)+... \ , \ \ \ \ \ \    \nu \ll \sqrt \S \ll 1  \ .    \la{zz}
  \eea
The fluctuation Lagrangian   will now have 4 of $S^5$ fields having mass $\nu^2$   and while
 the masses of the other fluctuation fields
become  \ci{ft1} (cf. \rf{lag},\rf{as},\rf{feq}):
\bea
&&\mu_t^2= 2 \rho'^2 -\kappa^2 +\nu^2,
 \qquad \mu^2_{\phi}=2 \rho'^2 -w^2 +\nu^2,
 \qquad \mu^2_{\rho}=2 \rho'^2 -w^2-\kappa^2+2 \nu^2, \no \\
 && \mu_{\beta}^2=2 \rho'^2 +\nu^2   \ , \qquad \mu_{\F} = \sqrt{ \rho'^2 +\nu^2} \ .
 \label{laes}
\eea
One  can then  compute the 1-loop correction to string energy by  expanding in the
 short string limit, i.e.  in $\eps \ll 1$
while  keeping  $\nu$  fixed.

Expanding the masses and  the coefficients in the mixing term in the fluctuation Lagrangian
we  get  the following expression for the 1-loop effective action (cf. \rf{abs}--\rf{avd})
\begin{eqnarray}
\Gamma_1(\eps^2)= &-&\frac{\T}{4 \pi}\int_{-\infty}^{\infty}d
\omega \bigg(8 \ln \frac{\det[\Delta_0 + \nu^2+
 \epsilon^2 \cos^2 \sigma]}{\det[\Delta_0+\nu^2 ]}-2
 \ln \frac{\det[\Delta_0 + \nu^2+ 2  \epsilon^2 \cos^2 \sigma]}{\det[\Delta_0 +\nu^2 ]}\nonumber\\
&-& \ln \frac{\det[Q_{\omega}]}{\det[Q_{\omega}^{(0)}]}+\ln \frac{\det[ P_{\omega}]}
{\det[Q_{\omega}^{(0)}]}\bigg) \ , \label{avdy}
\end{eqnarray}
where now
\begin{eqnarray}
P_{\omega}=\left(
             \begin{array}{ccc}
               -  \Delta_0 & 0 & 0 \\
               0 & \Delta_0 + \nu^2  & 0 \\
               0 & 0 &\Delta_0 + \nu^2  \\
             \end{array}
           \right) \ ,\  \ \ \ \ \ \ \ \ \ \ \ \ \  \Delta_0 \equiv - \partial_1^2 + \omega^2
	    \la{py}
\end{eqnarray}
and the  mixing term operator $Q_{\omega}$  is
 given to order $\eps^2$ by the following matrix ($i=1,2,3$)\foot{Here we expanded to order $\eps^2$
in small $\eps$ at fixed $\nu$ but in some terms formally  kept $\eps^2$ contributions  under
 the square roots to allow for  a  smooth $\nu\to 0$ limit.}
\bea
(Q_{\omega})_{1i}&=&\big\{ -(\Delta_0+\epsilon^2 \cos 2 \sigma);
 \ 0; \  2 \epsilon w \sin \sigma \sqrt{\nu^2 +\epsilon^2}\ \big\}\no \\
(Q_{\omega})_{2i}&=&\big\{0; \ \Delta_0 -1+
\epsilon^2 (\cos 2 \sigma+\ha); \ -2 \omega (1 + \ha \eps^2 \sin ^2 \s)
 \sqrt{\nu^2+1+\ha \epsilon^2}\big\}  \la{qqq}\\
(Q_{\omega})_{3i}&=&\big\{-2 \epsilon w \sin \sigma \sqrt{\nu^2 +\epsilon^2};
\ 2 \omega   (1 + \ha \eps^2 \sin ^2 \s)   \sqrt{\nu^2+1+\ha \epsilon^2 };
 \ \Delta_0 -1+
\epsilon^2 (\cos 2 \sigma-\ha)\big\}  \no
\eea
So far we considered
 $\epsilon \ll 1 $  with   $\nu$ arbitrary.
Next, we may specify either to the
 fast short string case ($\nu \gg \eps$)
or to the slow short  string case ($\nu \ll \eps$).
In the fast string  case
 we get
$Q_{\omega}=Q_{\omega}^{(0)}+ \epsilon Q_{\omega}^{(1)} + \epsilon^2 Q_{\omega}^{(2)}+...$
where
\begin{eqnarray}
Q_{\omega}^{(0)}=\bp
                -\Delta_0 & 0 & 0 \\
                 0 & \Delta_0 -1 & -2 \omega \sqrt{1+\nu^2} \\
                 0 & 2 \omega \sqrt{1+\nu^2} & \Delta_0 -1 \ep \ , \ \
Q_{\omega}^{(1)}=\bp
                 0 & 0 & 2 \omega \nu \sin \sigma \\
                 0 & 0 & 0 \\
                 -2 \omega \nu \sin \sigma & 0 & 0 \ep
		 \no \end{eqnarray}
\begin{eqnarray}
Q_{\omega}^{(2)}=\bp
                 -\cos 2 \sigma & 0 & 0 \\
                 0 & \cos 2 \sigma +\frac{1}{2} & - \frac{\omega}{ \sqrt{1+\nu^2}} [\ha +
		  (1+\nu^2)\sin^2 \sigma] \\
                 0 &  \frac{\omega}{ \sqrt{1+\nu^2}} [\ha +
		  (1+\nu^2)\sin^2 \sigma] & \cos 2 \sigma-\frac{1}{2} \ep \ .
\end{eqnarray}
We can  again diagonalize the  propagator matrix
 \begin{eqnarray}
D_{\omega}^{(0)}=M^{-1}(Q_{\omega}^{(0)})^{-1} M
= \left(
                \begin{array}{ccc}
                  -\frac{1}{n^2+\omega^2} & 0 & 0 \\
                  0 & \frac{1}{n^2+(\omega+ i \sqrt{1+\nu^2})^2 + \nu^2} & 0 \\
                  0 & 0 &  \frac{1}{n^2+(\omega - i \sqrt{1+\nu^2})^2 + \nu^2} \\
                \end{array}
              \right)\ ,
\end{eqnarray}
where $M$ is the  same as in \rf{my}. Similarly,
\begin{eqnarray}
&&D_{\omega}^{(1)}=M^{-1}Q_{\omega}^{(1)} M  = \bp
               0 & \omega \nu \sin \sigma & \omega \nu \sin \sigma \\
               - 2 \omega \nu \sin \sigma & 0 & 0 \\
               - 2 \omega \nu \sin \sigma & 0 & 0 \ep \ , \\
&&D_{\omega}^{(2)}=M^{-1}Q_{\omega}^{(2)} M  =\bp
               - \cos 2 \sigma & 0 & 0 \\
              0 & \cos 2 \sigma + \frac{i \omega [\ha + (1+\nu^2)\sin^2 \sigma]}{
	      \sqrt{1+\nu^2}} & -\frac{1}{2} \\
               0 & -\frac{1}{2} & \cos 2 \sigma - \frac{i \omega [\ha + (1+\nu^2)\sin^2 \sigma]}{
	        \sqrt{1+\nu^2}}\ep   \no
\end{eqnarray}
One can show that the last term in
\rf{avdy} vanishes. The leading term in  the short-string limit of $\G_1$
is of order  $\eps^2$.
To compute it we  note  that
\begin{equation}
\ln \frac{\det[-\partial_1^2+ \omega^2+\nu^2 + 2 \epsilon^2
\cos^2 \sigma]}{\det[-\partial_1^2+ \omega^2 +\nu^2 ]}\approx
\epsilon^2 \sum_n \frac{2\int_0^{2 \pi}\frac{d \sigma}{2 \pi}\cos^2 \sigma }{n^2+\omega^2+\nu^2}
 =\epsilon^2 \sum_n \frac{1}{n^2+\omega^2+\nu^2}. \label{lkl1}
\end{equation}
and use  the expansion
\begin{equation}
 \ln \frac{\det [A+ \epsilon B_1+\epsilon^2 B_2]}{\det A}=\epsilon \Tr[ A^{-1} B_1]+
  \epsilon^2 \Tr[A^{-1} B_2] -\frac{\epsilon^2}{2} \Tr[A^{-1} B_1 A^{-1} B_1] + ...
 \label{abg1}
\end{equation}
in the third  nontrivial term  in \rf{avdy}.
The order $\eps$ contribution vanishes. The $\eps^2$ terms come from
$\Tr[D_{\omega}^{(0)} D_{\omega}^{(2)}]$ and
$\Tr[D_{\omega}^{(0)} D_{\omega}^{(1)} D_{\omega}^{(0)} D_{\omega}^{(1)}]$.
Summing them up we get for the $\eps^2$ term in the effective action
\begin{eqnarray}
&&\Gamma_1(\eps^2) = \frac{\T\epsilon^2}{4 \pi} \int_{-\infty}^{\infty}
d\omega \sum_n \bigg(-\frac{2}{n^2+\omega^2+\nu^2}+\frac{ \nu^2+2 }{2 \sqrt{\nu^2+1}}\Big[
\frac{i \omega}{n^2+\nu^2+ (\omega+i \sqrt{\nu^2+1})^2} + c.c. \Big]
 \nonumber\\
&& \ \ \ \ \ \ \ \ \ \ \ \ \ \ \ \ \ \ \ \ \ \ -\ \frac{\nu^2 \omega^2}{2(n^2+\omega^2)}\Big[
\frac{1}{(n+1)^2+\nu^2+
(\omega+i \sqrt{\nu^2+1})^2} \nonumber\\
&& \ \ \ \ \ \ \ \ \ \ \  \ \ \ \ \ \ \ \ \ \ \ +\ \frac{1}{(n-1)^2+\nu^2+(\omega+i
\sqrt{\nu^2+1})^2} + c.c.\Big]\bigg)
\end{eqnarray}
Performing separate shifts of $\omega$  under the   integrals in  various terms
as discussed  in sections 3 and 4  gives
\begin{eqnarray}
&&\Gamma_1(\eps^2)= \frac{\T\epsilon^2 }{4 \pi} \int_{-\infty}^{\infty}
d\omega \sum_{n=-\infty}^{\infty} \nu^2 \bigg[\frac{1}{n^2+\omega^2+\nu^2}\nonumber\\
&& \ \ \  \ \ \ \ \ \ \ \ \ \ \ \ +
\ \frac{\omega^2 (n-1-\nu^2)-(\nu^2+1)(n+1 +\nu^2)}
{[(n+1)^2 +\nu^2 +\omega^2][(n+1 +\nu^2)^2+\omega^2 (\nu^2+1)]}\bigg] \ .
\end{eqnarray}
The  sum over $n$ can be performed
 exactly and we get
\begin{equation}
\Gamma_1(\eps^2) =  \frac{\T\epsilon^2}{4 \pi} \int_{-\infty}^{\infty}
d\omega\  \frac{\pi \sin (2 \pi \nu^2)}{\cos (2 \pi \nu^2) -\cosh (2 \pi
\omega \sqrt{\nu^2+1})}   \ ,
\end{equation}
or finally
\begin{equation}
\Gamma_1=\frac{\T \epsilon^2}{4}\frac{2 \nu^2-1}
{\sqrt{\nu^2+1}} + O(\eps^4) \ .
\end{equation}
Recalling that $E_1= \frac{\Gamma_1}{\kappa \T}$ and that in the
``short fast  string'' limit under the
 consideration (i.e. $\eps \ll 1, \ \epsilon \ll \nu$) one has
$\kappa= \nu+ \frac{\epsilon^2}{2 \nu}+...$, we finally obtain
\begin{equation}
E_1= \frac{\S}{2 \nu}\frac{2 \nu^2 -1}{\nu^2+1}  +   O(\S^2)   \ ,  \label{qad}
\end{equation}
where we have  replaced  $\epsilon$  by $\S$  using   (\ref{maq}).
So far $\nu$ here is arbitrary apart from the
condition $\nu \gg \eps $, i.e. $ 2 \S \ll \nu^2 \sqrt{1 + \nu^2}$,
so that  \rf{qad} is the 1-loop correction to the classical energy in
\rf{aml}.

Assuming further that   $\nu\gg 1 $ 
we get
\begin{equation}
E_1= \frac{\S}{\nu}   -\frac{3}{2}\frac{\S}{\nu^3}+...
=  { S \ov J} ( 1 -\frac{3}{2}\frac{\l}{J^2 }+...) + ...
 \ , \la{kt}
\end{equation}
which  should be  the correction to (\ref{jue}).

This expression may  be  compared to the 1-loop
correction to the folded spinning string
energy found by quantizing the $sl(2)$
 Landau-Lifshitz model in Appendix D of \ci{mtt}
 \be \la{hoh}
 E_1= -   { \S \ov 2 \nu^3} + O(\S^2)  = -
 { \l \ov 2 J^2 }{ S \ov J}  + O(\S^2) \ .\la{ops}
 \ee
 There one first have taken the  large $\nu$  limit with  $\S\ov \nu$  kept fixed
 and then expanded  in ${\S\ov \nu} \ll 1 $.
 Here the order of limits  was  different
 (we first expanded in $\eps$ for fixed $\nu$)
 and  that could be a possible reason for a  disagreement
  between \rf{kt} and \rf{ops}.\foot{The presence of an unusual
  $ {S \ov J}$  term in the 1-loop correction \rf{kt}
  may be an artifact of the limit of the above expansion procedure.}
To recover the standard fast string result one would need to start
with the short string fluctuation operators in (\ref{qqq}),
 where no assumption on $\frac{\S}{\nu}$ was made,
use them and (\ref{lkl1}) without expanding in $\epsilon$,
 compute the determinants needed in (\ref{avdy}),
 then expand in large $\nu$ with $\frac{\S}{\nu}$ kept fixed,
and at the end take $\frac{\S}{\nu}$ to be small.

\medskip 

One can then  consider the 1-loop correction in the small $\nu$  region by taking $\eps$ to zero
while  keeping the parameter $x\equiv
\frac{\nu}{\epsilon}$ fixed,  i.e. scaling $\nu$ to zero
together with $\eps$  so that ${\nu \ov \sqrt {2 \S}  } \approx x$
remains finite. 
We refer to the follow-up paper \ci{BT}   for the details.

\appendix
\subsection*{Appendix C:  A comment on regularization ambiguity }
\refstepcounter{section}
\def\theequation{C.\arabic{equation}}
\setcounter{equation}{0}

Let us start with the expression in the second line in (\ref{nad})
\begin{eqnarray}
X & \equiv &\sum_{n \neq 0, 1} \frac{\omega^2 (n-1)}{((n-1)^2+\omega^2)^2} -
\sum_{n \neq 0, -1} \frac{\omega^2 (n+1)}{((n+1)^2+\omega^2)^2}\nonumber\\
&=&\frac{2 \omega^2}{(\omega^2+1)^2}+\sum_n R_n(\omega) \ ,
\end{eqnarray}
where
\begin{equation}
R_n(\omega) =  \frac{\omega^2 (n-1)}{((n-1)^2+\omega^2)^2}- \frac{\omega^2 (n+1)}{((n+1)^2+\omega^2)^2}  \label{ar}
\end{equation}
We  need to compute $\int d \omega \sum_n X$ and thus
\be
Y\equiv \int^\infty_{-\infty} d \omega \sum^\infty_{n=-\infty}  R_n (\omega)  \ .
\la{hip} \ee
If we perform the sum over
$n$ first, as we did in the main text, then  result is zero, i.e.
\be   Y=0 \ . \ee
This can be seen right away, of course, by performing opposite $n$-shifts, i.e. $n \rightarrow n \pm 1$ in the two terms in $R_n(\omega)$.  Curiously, if  instead we perform the integral over $\omega$ first,
  and then do the sum we obtain
\begin{equation}
Y = -2 \pi \ .
\end{equation}

Looking in more detail at the origin of this  ambiguity one  discovers
 that it may be interpreted as  a UV regularization
 anomaly. Indeed, if  we replace
  the sum over $n$ by an integral, and introduce cutoffs
  $N$ and $L$  for the integral over  $n$ and $\omega$ respectively,
   we obtain
\bea
&&  Y \rightarrow Y(L,N)\equiv  \int_{-L}^L \int_{-N}^N d \omega d n \ R_n(\omega)
 \ , \\
&&Y(L,N)= 2\bigg[(N-1) \tan^{-1} \frac{L}{N-1}-(N+1) \tan^{-1} \frac{L}{N+1}\bigg]
 \ .
\eea
In accord with the above remarks we find
\be Y(L,\infty )= 0 \ , \ \ \ \ \ \
Y(\infty,N)= - 2 \pi  \ . \la{bu} \ee
More generally, we can take the
 limit $N,L \rightarrow \infty$ with $N = a L$ where
  $a$ is a fixed constant. Then we get a finite  result that depends on $a$
\begin{equation}
Y(a) \equiv  Y(L, a L)_{_{L\to \infty}} =
 \frac{4 a}{1+a^2}-4 \cot^{-1} a  \label{amu}  \ .
\end{equation}
The previous results in \rf{bu}  correspond to  the choice of $a=\infty $ or  $a= 0$:
The two limits are now
\be
Y(a=\infty) = 0 \ , \ \ \ \ \ \ \ Y(a=0) = -2 \pi  \ . \ee
In this
 paper we have chosen to  perform the sums first as this is
 is a natural prescription  to
  dealt with  the corresponding $2d$ functional determinants on $R \times S^1$.

In the absence of 2d Lorentz covariance (broken by our background and by the topology
 of the world sheet) it is not a priori clear
which regularization should  be preferred: that choice may be hidden in how one should
implement the global  symmetries of the superstring theory at the quantum level.
One possibility is to demand that since this regularization ambiguity
has UV nature, the regularization on
the world-sheet cylinder  $R \times S^1$ should be the same   as on $R^{1,1}$, i.e.
on the infinite plane   which appears  in the long string limit. That would
suggest that a UV cutoff should be imposed  in the 2d Lorentz-invariant way, i.e.
$\omega^2 + n^2  > \Lambda^2, \ \  \Lambda\to \infty$. Setting $\omega= p \cos \vp, \ \ n=p \sin \vp$
with $ p < \Lambda$
and integrating first over $\vp$ from 0 to $2 \pi$  we get  0
(assuming analytic continuation from relevant region of large $p$). Thus we  end up with
the same result as in the regularization we have preferred above in the main text.

\end{document}